\documentclass[fleqn,usenatbib]{mnras}

\usepackage[varvw]{newtxmath}
\usepackage{newtxtext}
\usepackage{appendix}
\usepackage{comment}
\usepackage{xcolor}

\usepackage[T1]{fontenc}

\DeclareRobustCommand{\VAN}[3]{#2}
\let\VANthebibliography\thebibliography
\def\thebibliography{\DeclareRobustCommand{\VAN}[3]{##3}\VANthebibliography}

\usepackage{graphicx}	
\usepackage{amsmath}	
\usepackage{tabularx}
\usepackage{orcidlink}

\defcitealias{Butler_2025}{B25}
\defcitealias{Martin_2026a}{M26}

\title[Assembly channels of ICL and DM]{Intracluster light and dark matter: shedding light on orbital differences through assembly channels}

\author[M. Butler et. al.]{
Madeline Butler\orcidlink{0009-0004-9273-4630}$^{1}$\thanks{E-mail: madeline.butler@nottingham.ac.uk},
Garreth Martin\orcidlink{0000-0003-2939-8668}$^{1}$,
Nina A. Hatch\orcidlink{0000-0001-5600-0534}$^{1}$,
Harley J. Brown\orcidlink{0009-0004-7521-8204}$^{1}$,
Yohan Dubois\orcidlink{0000-0003-0225-6387}$^{2}$,
\newauthor~Adela Fernandez\orcidlink{0009-0001-4673-526X}$^{1}$
and Frazer R. Pearce\orcidlink{0000-0002-2383-9250}$^{1}$
\\
$^{1}$School of Physics and Astronomy, University of Nottingham, University Park, Nottingham NG7 2RD, UK\\
$^{2}$Institut d’Astrophysique de Paris, UMR 7095, CNRS, Sorbonne Universit\'e, 98 bis boulevard Arago, 75014 Paris, France
}

\date{Accepted xxx. Received yyy; in original form zzz}

\pubyear{2026}

\begin{document}
\label{firstpage}
\pagerange{\pageref{firstpage}--\pageref{lastpage}}
\maketitle

\begin{abstract}
The intracluster light (ICL) is a promising tracer of the underlying dark matter (DM) distribution of galaxy clusters, but the intracluster stars are, on average, more tightly bound and have lower angular momentum than the DM. To identify the origin of this phase-space offset, we use galaxy clusters from the \textsc{Horizon-AGN} cosmological simulation to trace the stellar and DM components present in $z\approx0$ clusters back to their assembly channels. We find that material assembled at earlier times ($z\gtrsim 2$) occupies orbits with lower energy and angular momentum than late-assembled material, and is more centrally concentrated and morphologically rounder. However, within both early and late assembly channels, stars always occupy more bound orbits than the corresponding DM component. The offset therefore cannot be explained solely by differences in the mix of early- and late-assembled material, but instead arises independently within each channel and is preserved as the cluster grows. The differential tidal stripping of stars and DM is strongest for material accreted during more equal-mass mergers at earlier times: the DM-dominated outskirts of satellites are stripped first, on higher-energy orbits, whereas the stars are only stripped once the satellites have sunk to lower-energy orbits via dynamical friction. A substantial fraction of late-assembled DM arrives through smooth accretion -- a channel with no stellar counterpart. This material remains on high-energy orbits and contributes significantly to the offset. Phase-space offsets between the ICL and DM are therefore a predictable consequence of hierarchical cluster assembly and encode the dynamical history through which clusters grow.
\end{abstract}




\section{Introduction}\label{sec:introduction}

In the standard model of structure formation in a $\Lambda$CDM Universe, objects assemble over cosmic time through hierarchical merger events and gradual accretion of surrounding material. Galaxy clusters are the most massive virialised objects predicted to form by the current epoch of the Universe, typically defined by their halo mass exceeding $10^{14}\,\textrm{M}_\odot$. They are important from both a cosmological and astrophysical perspective: their integrated properties across the population (e.g. masses, number densities and scaling relations) are used to constrain cosmological parameters \citep[see][for a review]{Allen_2011, Kravtsov_2012}, whilst their resolved properties (e.g. mass and luminosity distributions) can probe the particle nature of dark matter \citep[DM;][]{Robertson_2018, Brinckmann_2018, Harvey_2019, Montes_Trujillo_2019} and galaxy evolution processes \citep{Peng_2010, Bahé_2015, Carleton_2019}.

Although the mass budgets and consequently also the dynamics of galaxy clusters are dominated by their DM halo \citep{Gonzalez_2007, Gonzalez_2013}, the distribution of baryonic content serves a crucial function in providing observable tracers of the gravitational potential. One such established tracer is the intracluster medium (ICM), which contains the majority of the baryons and is detectable in X-rays emitted via thermal bremsstrahlung \citep{Borgani_2001, Ettori_2013}. Another tracer is the dynamics of satellite galaxies \citep[e.g.][]{Gifford_2013}, collectively composing a significant fraction of the cluster stellar material. The remaining stars reside either in the brightest cluster galaxy (BCG, typically a massive quiescent galaxy occupying the bottom of the cluster potential well) or within the intracluster space. The combined emitted radiation of these intracluster stars is referred to as the intracluster light \citep[ICL; see][for a review]{Montes_2022}, a diffuse low-surface-brightness (LSB) component of the cluster, although this term is often used as shorthand to refer to the intracluster stars themselves.

The ICL typically comprises 10 to 20 per cent of the stellar mass of the cluster, although values in the literature range from 5 to 40 per cent \citep[e.g.][and references therein]{Brough_2024, Zhang_2024, Brown_2026, Gully_2026}, due to both genuine scatter between clusters and varying methodologies to extract the ICL. The primary formation channel of the ICL is through the liberation of stellar material from infalling galaxies throughout the cluster assembly. This occurs via a combination of tidal stripping \citep{Rudick_2009, Contini_2014, Brown_2024, Martin_2024, Golden-Marx_2025} and ejection during violent relaxation processes \citep[usually between massive infallers and the BCG;][]{Willman_2004, Murante_2007, Conroy_2007, Groenewald_2017, Contini_2018}. A considerable contribution of "pre-processed" ICL stars, from the pre-existing diffuse stellar component of infalling groups and clusters, is also expected in more massive host clusters \citep{Rudick_2006, Mihos_2017, Chun_2023, Chun_2024, Joo_Jee_2023, Contini_2024}.

The study of ICL has gained significant traction in the past decade. On the theoretical side, numerous cosmological hydrodynamic simulations with a range of numerical methods and treatments of astrophysical processes (such as gas heating and cooling, star formation, AGN feedback, etc.) now provide large samples of galaxy clusters with self-consistent ICL production \citep[see][for inter-simulation comparison studies]{Kimmig_2025, Brown_2026, Martin_2026a}. Importantly, the DM mass resolution is generally high enough in these modern simulations to achieve sensible tidal stripping rates \citep{Martin_2024}. Applying standardised ICL definitions to a suite of simulations, \cite{Brown_2026} found the bulk properties of the ICL (stellar mass fraction, main progenitor objects, redshift evolution) are approximately converged. On the observational side, the advent of current and upcoming facilities, namely the European Space Agency's \textit{Euclid} \citep{Laureijs_2011, Euclid_Collaboration_Mellier_2025} Wide Survey \citep[EWS;][]{Scaramella_2022} and the Vera C. Rubin Observatory's Legacy Survey of Space and Time \citep[LSST;][]{Ivezic_2019}, have begun revolutionising LSB science \citep[e.g.][]{Martin_2022, Euclid_Collaboration_Bellhouse_2025} by providing deep, wide imaging of thousands of clusters across a range of masses and redshifts. In combination with careful imaging processing techniques \citep[][]{Bamford_2026}, systematic resolved studies of the ICL in multiple photometric bands are now possible.

The ICL has also been proposed as a possible tracer of the underlying DM halo of the cluster \citep{Montes_Trujillo_2019}. The basic theoretical motivation for an expected connection between the ICL and DM can be understood as follows. First, a significant fraction of both the DM and intracluster stars share a common origin in infalling galaxies, groups and smaller clusters. Second, in the context of galaxy cluster stellar densities, the intracluster stars can be treated as collisionless, and thus are subject to the same physics as the DM (which is collisionless in the cold DM model by construction). The intracluster stars and DM therefore retain memory of their progenitor object merger properties at the time of their unbinding, which should persist for timescales far exceeding that of a Hubble time assuming sufficiently adiabatic cluster evolution \citep[a reasonable assumption for cluster-scale haloes at late times;][]{Wechsler_2002, Zhao_2003, McBride_2009}. The existence of LSB tidal features such as streams and shells are clear evidence of this on $\sim$\,Gyr timescales \citep[e.g.][]{Mihos_2004, Rudick_2009, Mihos_2017, Pop_2018, Martin_2022, Martínez-Delgado_2023, Khalid_2024}, but properties such as orbital energy and angular momentum still remain approximately constant \citep{Binney_Tremaine_2008} even after these obvious merger signatures disappear due to phase-mixing. The collisionless nature of the intracluster stars sets the ICL apart from the (plasmatic) hot ICM as a potential DM tracer; not only can the ICM thermalise over $\sim$\,Gyr timescales (with the orbital energy and angular momentum exchange between particles quickly erasing the memory of their original orbits), but it is also subject to additional baryonic physics. Most relevant for clusters is the injection of additional energy and angular momentum from AGN outflows and jets which can redistribute the ICM \citep{Giacintucci_2011, Planelles_2014, Hlavacek-Larrondo_2015}, and the "sloshing" that occurs during major mergers due to high-velocity accretion of the merging object's own ICM \citep[e.g. the bullet cluster;][]{Clowe_2004, Markevitch_2004}.

One aspect of the ICL-DM correspondence that has been explored is the morphological shape. By comparing the total mass distribution inferred through strong gravitational lensing to the stellar light distribution of the combined BCG+ICL component, \cite{Montes_Trujillo_2019} found that the surface brightness contours of the ICL matched closely with the total mass contours in the Hubble Frontier Fields clusters, often providing a closer match than ICM contours as traced by X-ray emission. This correspondence between the ICL and total mass has been reproduced in other observational studies \citep{Kluge_2021, Diego_2023, Diego_2024, Cha_2025, Ellien_2025} and between the ICL and DM in hydrodynamic simulations \citep{Alonso_Asensio_2020, Gonzalez_2021, Yoo_2022, Yoo_2024, Yoo_2025, Reina-Campos_2023, Fernandez_2026, Martin_2026b}, using different methodologies of measuring the shape similarity such as the weighted overlap coefficient \citep[WOC;][]{Yoo_2022} and comparing the parameters of the moment-of-inertia or shape tensors \citep{Zemp_2011}.

However, these findings must be interpreted with caution. Morphological similarity alone does not necessarily imply a similar relationship between other integrated properties of the intracluster stars and DM. In particular, it is tempting to construe the similarity in morphological shapes as the intracluster stars simply tracing the underlying gravitational potential generated primarily by the DM. Instead, \cite{Martin_2026b} showed that the ICL and DM inherit this close correspondence in morphological shape from their shared accretion geometry. Nor does a close match in the morphological shapes of the ICL and DM imply a corresponding match in mass distribution. Both theoretical \citep{Pillepich_2018, Alonso_Asensio_2020, Chen_2022, Reina-Campos_2023, Contreras_Santos_2024, Butler_2025, Manuwal_2025, Jeon_2026} and observational \citep{Sampaio_Santos_2020, Diego_2023} literature has established that the ICL is more centrally-concentrated than the DM -- or, in other words, ICL radial density profiles are much steeper than that of their DM counterparts, such that their ratio approximately follows a power law with a negative slope. Additionally, the kinematics of the ICL are systematically lower than the DM \citep{Dolag_2010, Contreras_Santos_2024, Butler_2025}. In our previous work \cite[][hereafter \citetalias{Butler_2025}]{Butler_2025}, we connected these differences back to the underlying orbital properties, demonstrating that the intracluster stars tended to be on lower energy and more anisotropic (lower angular momentum) orbits than the DM. This discrepancy was still sensitive to the ICL definition employed, but importantly could not be reconciled by any sensible BCG-ICL separation.

The emerging explanation of these systematic offsets in phase space involves a combination of two effects. First, the DM halo of infalling haloes is already established as being more spatially extended and more loosely-bound on average than its stellar component, thus is more vulnerable to efficient tidal stripping \citep{Villalobos_2012, Smith_2016b, Joshi_2019, Haggar_2021, Martin_2024}. Second, massive satellites (with stellar mass larger than that of the Milky Way) which represent the dominant progenitor objects of the ICL \citep[e.g.][]{DeMaio_2018, Montes_Trujillo_2018, Montes_2021, Chun_2023, Chun_2024, Ahvazi_2024b, Brown_2024, Brown_2026} are subject to significant orbital decay upon entering the cluster \citep{Tormen1997}, through processes such as dynamical friction \citep{Chandrasekhar1943} and tidal torques. The proposed mechanism posits that the majority of the DM halo is stripped rapidly upon cluster infall, retaining its initial orbital properties, whilst the stars remain bound until later times, when the merging halo has sunk to a lower energy and angular momentum orbit, and the tidal forces acting on the merging halo are eventually sufficient to liberate this stellar component.

In \cite{Martin_2026a}, hereafter \citetalias{Martin_2026a}, we demonstrated in controlled merger simulations that this mechanism can produce a systematic offset between the ICL and DM orbital properties, comparable to that seen in the cosmological simulations when convolved with realistic satellite infall populations. However, a full cosmological context is required to understand whether this mechanism acts as predicted, and can solely explain the phase space differences between the ICL and DM in realised instances of simulated clusters. First, mergers are not the only cluster mass assembly channel; the ability for further star formation to occur within the cluster and the presence of DM outside of collapsed structures must be accounted for. Second, the timing of mergers (and other assembly channels) is important to consider for not only their initial orbits, but also the progression of tidal stripping by redshift $z=0$. Third, any interactions between infalling haloes, either directly or through non-adiabatic evolution of the cluster itself, may add complexity.

This work explores the origin of orbital, spatial and morphological differences between the intracluster stars and DM through categorising the $z=0$ content of simulated galaxy clusters by its channel of assembly. In Section \ref{sec:simulation_and_method} we describe the key details of the \textsc{Horizon-AGN} simulation, outline our methodology to identify structures and mergers, and how we categorise material into assembly channels. In Section \ref{sec:results_assembly_channels} we investigate the mass fractions and average orbital properties of the material in each assembly channel, and in Section \ref{sec:results_halo_accretion} we explore mergers grouped by merger time and mass ratio. We then assess the radial profiles and morphological shapes of the BCG+ICL stars and main halo DM in each assembly channel in Section \ref{sec:results_spatial_distributions}. We then discuss the insights provided by our analysis in Section \ref{sec:discussion}, before summarising our conclusions in Section \ref{sec:conclusions}. Unless otherwise stated, all measurements are quoted in proper units and do not contain additional factors of $h\equiv H_0/100\textrm{kms}^{-1}\textrm{Mpc}^{-1}$.

\section{Simulation \& method}\label{sec:simulation_and_method}

\subsection{\textsc{Horizon-AGN}}\label{sec:horizon_agn}

In this work, we use the \textsc{Horizon-AGN} cosmological hydrodynamic simulation \citep{Dubois_2014,Dubois_2016}. \textsc{Horizon-AGN} adopts a standard $\Lambda$CDM cosmology with the following parameters: $\Omega_\textrm{m}= 0.272$, $\Omega_\Lambda= 0.728$, $\Omega_\textrm{baryon}= 0.045$, $\sigma_8 = 0.81$, $H_\textrm{0} = 70.4\,\textrm{km}\,\textrm{s}^{-1}\,\textrm{Mpc}^{-1}$ and $n_\textrm{s}=0.967$. These parameters are compatible with the \textit{WMAP7} cosmology \citep{Komatsu_2011}, and also with \cite{Planck_2014} within a 10 per cent relative variation. The simulation box with side length $100\,h^{-1}\,\textrm{comoving Mpc}$ contains $1024^3$ DM particles with a mass resolution of $M_\textrm{DM,res} = 8\times10^7\,\textrm{M}_\odot$, and an initial gas resolution of $M_\textrm{gas,res} = 1\times10^7\,\textrm{M}_\odot$. The N-body and hydrodynamical solver used is the adaptive mesh refinement code \textsc{Ramses} \citep{Teyssier_2002}. The initial mesh begins with a refinement of level ten, with up to four additional levels of refinement down to a spatial resolution of $\Delta x = 1\,\textrm{kpc}$. The refinement is performed in a quasi-Lagrangian manner, triggered by the total baryonic or DM mass in a cell exceeding $8\,M_\textrm{DM,res}$, and an additional level of maximum refinement is introduced each time the expansion scale factor doubles (first at $a=0.1$) to keep $\Delta x$ approximately constant in physical units. The simulation begins at $z\approx38$, with the initial conditions generated using {\sc mpgrafic} \citep{Prunet_2008}. The particle data is saved with a time resolution of $\approx250\,\textrm{Myr}$ up to $z=0.018$. The most recent simulation snapshot used for this work, $z=0.056$, will be referred to as the $z\approx0$ snapshot.

Gas heating from a uniform UV background takes place from $z=10$ \citep[following][]{Haardt_1996}. Gas cooling by H, He and a contribution from metals down to $10^4$\,K is modelled using a \citep{Sutherland_1993} scheme. Star formation occurs when the threshold number density of H exceeds $0.1\,\textrm{cm}^3$ according to a Schmidt law \citep{Schmidt_1959} with star formation rate density $\dot{\rho}_\star = \varepsilon_\star\rho/t_\textrm{ff}$. Here, $\varepsilon_\star = 0.02$ \citep{Kennicutt_1959, Krumholz_2007} and $t_\textrm{ff}$ is the local free-fall time of the gas. This produces stellar particles with initial $M_\textrm{star,res} \approx 2\times10^6\,\textrm{M}_\odot$ following a \cite{Salpeter_1955} initial mass function with lower and upper mass limits of $0.1\,\textrm{M}_\odot$ and $100\,\textrm{M}_\odot$ respectively. Further details of star formation and the implementations of stellar and active galactic nuclei feedback can be found in \cite{Kaviraj_2017}.

{\sc Horizon-AGN} and similar large-volume cosmological hydrodynamical simulations have been extensively utilised for studies of the resolved ICL (e.g. \citealt{Cañas_2020, Brough_2024, Brown_2024, Brown_2026}, \citetalias{Butler_2025}, \citealt{Kimmig_2025}, \citetalias{Martin_2026a}). One important reason for this is the combination of mass resolution and simulated volume, which is required for sensible tidal stripping rates \citep{Martin_2024} and thus ICL production, whilst still simulating a sufficiently large volume to allow the formation of a reasonable number of cluster-mass haloes. {\sc Horizon-AGN}'s mass resolution and cluster sample size is comparable to other cosmological simulations of similar volume (e.g. \textsc{TNG100}, \citealt{Pillepich_2018}) and cluster zoom-in simulations (e.g. \textsc{Hydrangea}, \citealt{Bahe_2017}).

{\sc Horizon-AGN} is not calibrated to the local Universe, except from choosing black hole (BH) feedback parameters to match observed BH mass relations at $z=0$. Aggregate star formation histories of the model galaxies broadly reproduce that of galaxies in the real Universe \citep{Kaviraj_2017} using observational data in the redshift range $0<z<6$, showing good agreement of luminosity functions, stellar mass functions, the star formation main sequence, the cosmic star formation history and rest-frame ultraviolet--optical--near infrared colours. 

Compared to \textsc{TNG100} and \textsc{Hydrangea}, \citet{Brown_2026} reported slightly fewer low-mass satellites (stellar mass $M_{\star}\lesssim10^{10}$\,M\textsubscript{\sun}) joining to-be $z\approx0$ \textsc{Horizon-AGN} clusters between $z\sim0$ and $z\sim2$, with a trace excess of accreted $M_{\star}\sim10^{11}$\,M\textsubscript{\sun} satellites also noted -- speculated by those authors to jointly stem from the supernova feedback implementation of \textsc{Horizon-AGN} possibly being slightly understrength (also see \citealt{Kaviraj_2017}). Additionally, \citet{van_de_Sande_2019} noted the size-mass relation for $M_{\star}\gtrsim10^{10}$\,M\textsubscript{\sun} galaxies in \textsc{Horizon-AGN} to be offset towards larger radii as compared to \textsc{Eagle} \citep{Crain_2015, Schaye_2015}, \textsc{Hydrangea}, and \textsc{Magneticum} \citep{Dolag_2025} galaxies as well as observations (see \citealt{Peirani_2019}, \citealt{Martin_2019}, and \citealt{Brown_2026} for further relevant discussion). 

\subsection{Structures at $z\approx0$}\label{sec:method}

\begin{figure*}
	\includegraphics[width=\textwidth]{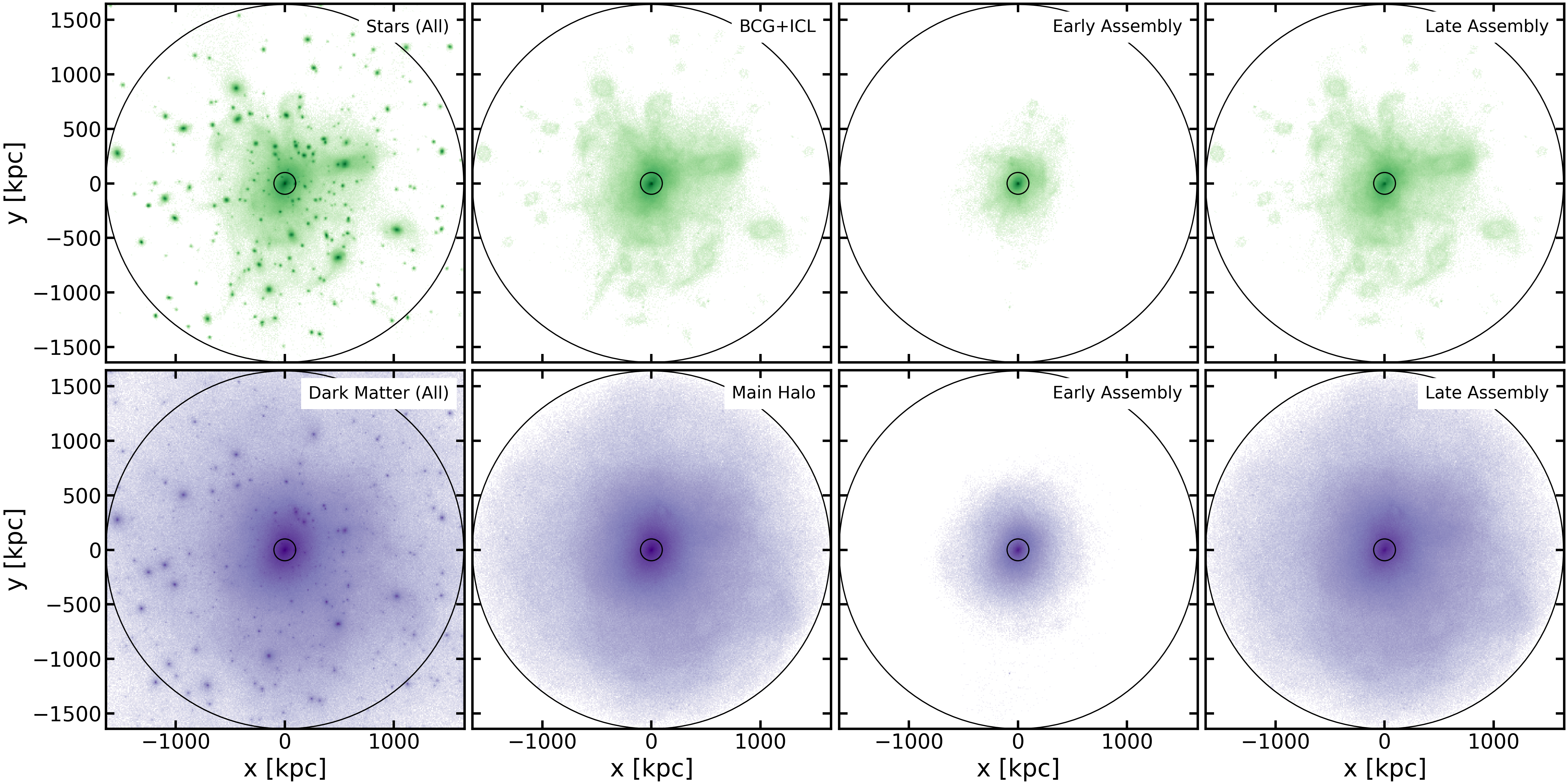}
    \caption{Projected surface density map of the BCG+ICL stars (top, green) and main halo DM (bottom, purple) along two arbitrary spatial axes centred on the main halo for Hrz049 at ${z\approx0}$. In all panels, darker colours represent higher projected surface densities on a logarithmic scale. The top panels show, from left to right: all star particles in a cube of volume ($R_{178}$)$^3$, the BCG+ICL material, BCG+ICL stars categorised as "early assembly", and BCG+ICL stars categorised as "late assembly". The bottom panels show, from left to right: all DM particles in a cube of volume ($R_{178}$)$^3$, the main halo material, DM categorised as "early assembly", and DM categorised as "late assembly". The inner and outer black circles indicate 100\,kpc and $R_{178}$ around the centre respectively. The colour normalisation of the early and late assembly panels are set by the BCG+ICL or main halo panels.}
    \label{fig:z0_map}
\end{figure*}

DM haloes are identified using the \textsc{AdaptaHOP} structure finder \citep{Aubert_2004} applied to the DM particles, updated by \cite{Tweed_2009} to ensure substructures always have lower mass then their host (sub)haloes. First, the local density associated with each particle is computed using the 20 nearest neighbours. Particles with a density above the threshold of 80 times the total matter density $\rho_\textrm{M}$ of the entire simulation box are then linked with their closest local maxima via a density walk. Groups of particles containing at least 100 particles are then defined as structures, resulting in a minimum detectable DM (sub)halo mass of $8\times10^9\,\textrm{M}_\odot$. Galaxies are identified in a similar way independently from the DM haloes, with the \textsc{AdaptaHOP} structure finder applied to the star particles. However, the density threshold is instead $178\,\rho_\textrm{M}$, and the minimum number of particles in a (sub)structure is 50. This corresponds to a minimum detectable galaxy mass of $\sim\,10^8\,\textrm{M}_\odot$. Galaxies are then associated with DM haloes by linking the most massive galaxy within $0.1\,R_\textrm{178, halo}$ of each DM halo as the central galaxy. Here, $R_\textrm{178, halo}$ is the radius of a sphere around the halo centre with average DM density $\bar{\rho}_\textrm{DM} = 178\,\rho_\textrm{M}$, computed by \textsc{AdaptaHOP} using the distribution of DM particles assigned to the structure, exclusive of substructure. $M_\textrm{178, halo}$ is the corresponding total mass of these DM particles within $R_\textrm{178, halo}$.

There are 14 haloes with masses $M_\textrm{178, halo}\geq 10^{14}\,\textrm{M}_\odot$ in \textsc{Horizon-AGN} at $z\approx0$, which we identify as cluster haloes. Hereafter, $R_\textrm{178}$ and $M_\textrm{178}$ are used to refer to the $R_\textrm{178, halo}$ and $M_\textrm{178, halo}$ of these cluster haloes respectively. We remove 3 of these clusters for breaks in the main progenitor branches of their merger trees (see Section\,\ref{sec:halo_tracking}), with the remaining 11 clusters being chosen as our sample. This is identical to the sample used in \cite{Brown_2024}, and has significant overlap with the samples used by \citetalias{Butler_2025} and \cite{Brown_2026}.

\subsubsection{Defining the BCG, ICL and satellites}

Before investigating the assembly of the stellar and DM content of each cluster, it is informative to categorise the star and DM particles into different components based on their spatial positions in the $z\approx 0$ cluster. In practice, this means attributing particles to substructure (satellite galaxies and DM subhaloes), and then any remaining particles within a sphere of radius $R_\textrm{178}$ around the BCG centre\footnote{The "BCG centre" here is the centre of the galaxy linked with main halo, which is typically within $\sim10\,\textrm{kpc}$ of the main halo centre.} are assigned to the combined BCG+ICL component (or the "main halo" component for DM). We employ the same definitions used in \citetalias{Butler_2025} to identify substructure (see their section 3.1 for full detail and justification of these choices). In brief, DM subhaloes are simply those found by \textsc{AdaptaHOP} with no modifications. The satellite galaxies on the other hand have an additional spherical mask applied, with the radius of this sphere related to the typical half-stellar mass radius of galaxies found in the $z\approx0$ cluster sample. The resulting BCG+ICL component therefore is relatively conservative, only containing stars which are sufficiently far away from any satellite galaxy.

There is no straightforward way to separate the ICL from the BCG, given that both components (broadly) share a common dominant origin in merger events and both their centres should approximately coincide with the bottom of the cluster potential well. We therefore often use the combined BCG+ICL component and compare to the DM in the main halo throughout this paper, which represents the material that has been processed out of satellite galaxies by $z\approx0$. When separating these components is required, we define the BCG as all star particles within a sphere of radius 100\,kpc, and the ICL as a spherical shell with inner radius 100\,kpc and outer radius $R_\textrm{178}$. The equivalent definitions for the DM are the "BCG region" and "ICL region" respectively. Comparisons between the stellar and DM components are always made in the equivalent regions, thus for readability we omit labelling the equivalent DM region in the text where it is unnecessary. An example of a projected density surface map can be seen in the left half of Fig.~\ref{fig:z0_map}, showing the isolated BCG+ICL (top) and main halo (bottom) component. 

The choice of a 100\,kpc aperture cut is relatively conservative for the ICL, at the upper end of the typically-quoted values for the transition radius of 30\,kpc (e.g. in \textsc{TNG100}) to 100\,kpc, where the density profile of the ICL becomes dominant over the BCG \citep{Contini_2022, Brough_2024, Zhang_2024, Golden-Marx_2025, Kluge_2025}. \cite{Pearce_2026} investigate the effectiveness of a fixed central aperture by comparing it with a BCG--ICL classification based instead on the assembly history and subsequent dynamical fate of the stellar material. The radial profiles obtained from the dynamical classification show the BCG contribution becoming subdominant at radii of order $0.04-0.05\,R_{200}$, corresponding to a characteristic scale of order $100\,\mathrm{kpc}$ for massive clusters. These results suggest that a fixed aperture of order $100\,\mathrm{kpc}$ provides a reasonable proxy for the central galaxy, while recognising that no sharp physical BCG--ICL boundary exists.

\subsubsection{Mass fractions}

It is important to quantify the relative significance (i.e. mass fraction) of the satellites, BCG and ICL at $z\approx0$. This is because understanding the selection effects imposed when masking the satellites to study the BCG+ICL component, then further masking the BCG to isolate the ICL (or the equivalent BCG region and ICL region for DM) gives insights into the bulk differences between stellar and DM components. We note that the percentages displayed are the mean between clusters, with a large inter-cluster scatter expected due to the range of assembly histories and current dynamical states.

The right-hand side of Fig.~\ref{fig:mass_contributions} gives the mass fractions of the satellites, BCG and ICL for the stellar material (top), and the satellites, BCG region and ICL region for the DM material (bottom). We find that 56.5 per cent of stars remain in satellites by $z\approx0$, whereas only 10.8 per cent of the DM is in $z\approx0$ satellites. This satellite stellar mass fraction is corroborated by \cite{Brown_2024}, who arrive at a very similar value (55 per cent) despite employing different methodology. This is somewhat unsurprising, given the large overlap in the cluster sample, but they do use a different method of separating satellite galaxies from the BCG+ICL. \citealt{Brough_2024}, \citealt{Kimmig_2025} and \citealt{Brown_2026} find differences in the median satellite fractions between simulations, so we do not compare to other simulations here, but note that the satellite mass fractions measured for the \textsc{Horizon-AGN} clusters in these studies are broadly consistent with our own results.

The diminished abundance of DM in satellites at $z\approx0$ compared to their stellar equivalent can be attributed to both the increased efficiency of tidal stripping (due to the DM being, on average, much more loosely bound than the stars in infalling haloes) and the existence of the smooth accretion assembly channel (discussed further in Section\,\ref{sec:late_assembly_dm}). A DM substructure fraction of $\sim10$ per cent in cluster-scale haloes is well-established in CDM simulation literature \citep{Ghigna_1998, Ghigna_2000, Springel_2001}, albeit with significant halo-to-halo variation due to e.g. halo concentration and environment \citep{Gao_2004, Gao_2011, Contini_2012, Hunde_2025}.

The remaining $\sim45$ per cent of stars and $\sim90$ per cent of DM belongs to the BCG+ICL and main halo components respectively. For the stars, this is almost evenly split between the BCG and ICL, while for the DM the ICL region is supremely dominant over the BCG region. This reflects the more centrally-concentrated nature of the stellar component compared to the DM component. Importantly, this also suggests that the bulk properties of the DM component are likely insensitive to the masking of the satellites and BCG region, unlike the stars where the equivalent masking removes three quarters of the stellar mass.

\subsection{Halo tracking}\label{sec:halo_tracking}

\subsubsection{Generating merger trees}\label{sec:merger_trees}

We are interested in quantifying the amount of material contributed to clusters by different assembly channels (e.g. mergers, smooth accretion, etc.), which requires robust tracking of the \textsc{AdaptaHOP} structures between snapshots by generating merger trees. However, as previously described, \textsc{AdaptaHOP} is applied independently on the stellar and DM distributions, and the resulting structures are only associated afterwards. Likewise, this results in two independent sets of merger trees, with no guarantee that a given stellar structure will remain linked with the same DM structure during its lifetime, and vice-versa. In other words, the main progenitor of the $z\approx0$ cluster halo is not necessarily linked with the main progenitor of the $z\approx0$ BCG in earlier snapshots, and the same logic applies to any halo/galaxy in the simulation.

For simplicity, we instead exclusively track DM haloes, and assign any star particles within $R_\textrm{178, halo}$ to these haloes at each snapshot to create unified structures. If a star particle is within $R_\textrm{178, halo}$ of multiple haloes (which is almost always the case when a star particle belongs to substructure), it is assigned to the halo with the smallest $R_\textrm{178, halo}$ to ensure substructure is accounted for.

\begin{table*}
\caption{Summary of the assignment of stellar and DM material to assembly channels. Halo accretion and smooth accretion are sub-channels of the late assembly channel. Full definitions are detailed in Section \ref{sec:assembly_channels}.}
\label{tab:definitions}
\begin{tabularx}{\textwidth}{|c|X|X|}
\hline
Assembly Channel & Definition for stars & Definition for DM \\
\hline
Early Assembly & Stars assembled into the cluster before $z\approx2$. Also includes material formed in-situ in the original cluster subhaloes after $z\approx2$, and material formed in-situ directly in the BCG+ICL after $z\approx2$. & DM assembled into the cluster before $z\approx2$. \\
\hline
Late Assembly & Stars assembled into the cluster after $z\approx2$. & DM assembled into the cluster after $z\approx2$. Split into the halo accretion and smooth accretion sub-channels. \\
\hline
Halo Accretion & The only sub-channel of the late assembly channel for stars. Includes material identified in mergers, material formed in-situ in subhaloes accreted after $z\approx2$, and material not assigned to other channels. & DM identified as part of mergers. \\
\hline
Smooth Accretion & Sub-channel not present for stars. & DM not identified as part of mergers. Represents material in unresolved merging haloes, material only weakly/not bound to merging haloes, and material in-fallen as part of larger cosmic structures. \\
\hline
\end{tabularx}
\end{table*}

We generate halo merger trees from the snapshots using the {\sc TreeMaker} algorithm \citep[originally developed by][]{Hatton_2003}. This scheme identifies descendants of a structure by the maximum mass overlap between the structure in a given snapshot and structures in the subsequent snapshot, and similarly identifies main progenitors of a structure by the maximum mass overlap between the structure in a given snapshot and structures in the previous snapshot. The updated version \citep{Tweed_2009} expands this scheme to include substructures in the merger trees. We impose that every cluster halo in our sample must have an unbroken main progenitor branch back to the chosen earliest snapshot. Extreme merging scenarios can cause the main progenitor branch of merger trees to break \citep{Srisawat_2013}. This occurs for clusters Hrz078, Hrz137 and Hrz157, and thus these clusters are removed, reducing the sample from the 14 to 11 clusters. Prior to $z\approx2$, the progenitors of $z\approx0$ clusters are often undergoing frequent mergers with similar-mass haloes, simultaneously making robust merger trees more difficult to construct at early times, and choosing just one halo branch to follow less physically meaningful. For these reasons, the earliest snapshot used is $z=1.97$, which will be referred to as the $z\approx2$ snapshot. We therefore have 40 snapshots at intervals of $\sim250\,\textrm{Myr}$ from $z=1.97$ to $z=0.056$, and extract cubes with side lengths of 8 comoving Mpc centred on the main halo progenitor for each snapshot.

\subsubsection{Merger tree corrections}

We manually check each main halo progenitor branch to ensure the position, velocity, $M_\textrm{178}$ and $R_\textrm{178}$ of the cluster are sensible, and correct a few instances where the halo position reported by \textsc{AdaptaHOP} is calculated incorrectly, which also affects the $M_\textrm{178}$ and $R_\textrm{178}$ values reported. We do this using the the centre-of-mass position of the particles within a 10\,kpc sphere of the halo centre in the previous snapshot, and interpolating between the $M_\textrm{178}$ and $R_\textrm{178}$ of the adjacent snapshots. There are also some instances of temporarily tracking substructure within the cluster as the main halo centre, which is corrected by using the properties of the "host" halo. Notably, in one instance (Hrz019), the cluster undergoes a merger event involving four similar-mass haloes at $z\approx1.5$, but is tracking the wrong halo prior to this event. Once these corrections have been made, all of the main halo progenitor properties evolve smoothly between snapshots.

\subsection{Assembly channels}\label{sec:assembly_channels}

\begin{figure*}
	\includegraphics[width=0.95\textwidth]{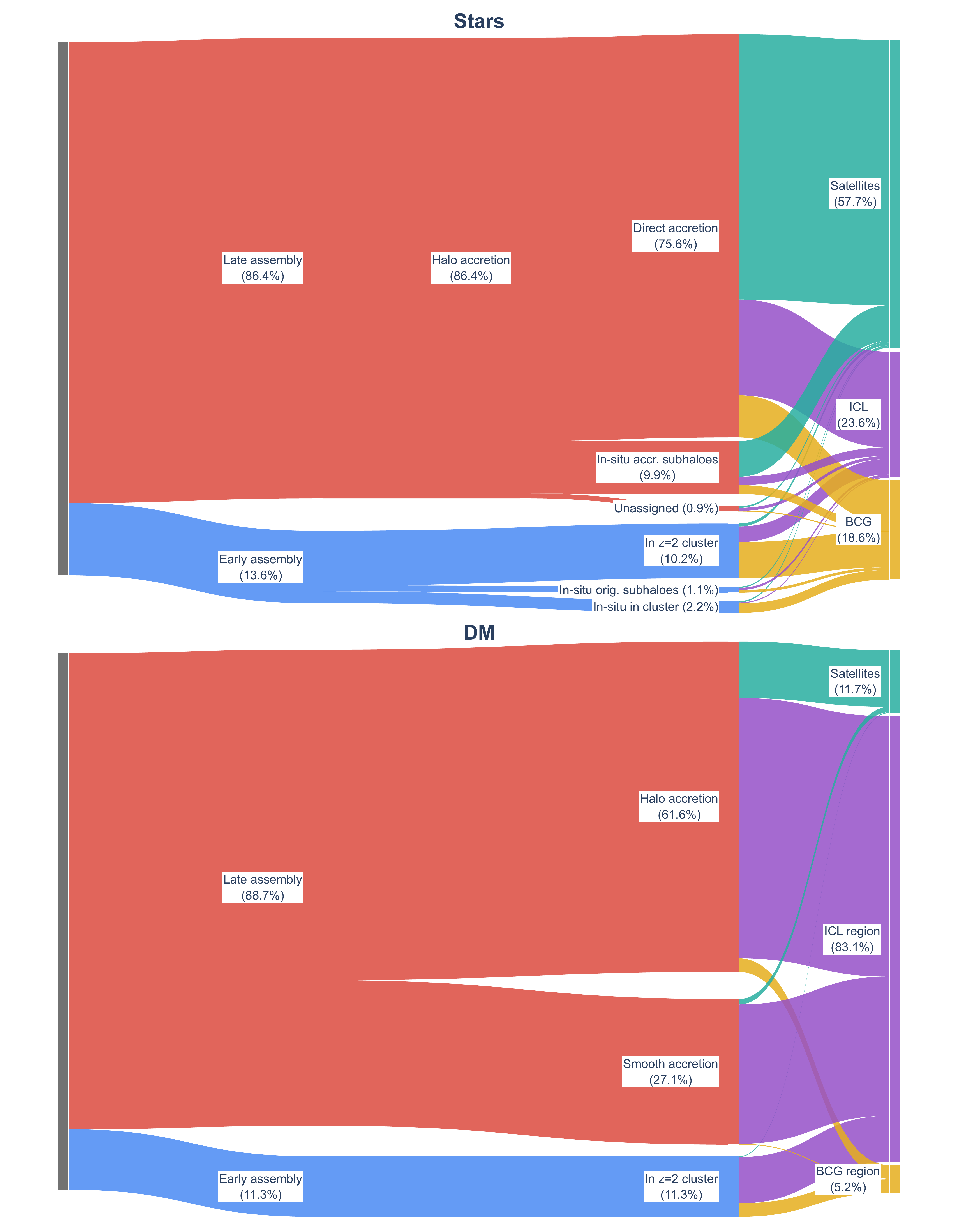}
    \caption{The mass contribution of each assembly (sub-)channel to the $z\approx0$ stellar (top) and DM (bottom) material. The percentages are relative to the total stellar or DM content, and the mean between clusters is shown. Each link width is proportional to the percentage mass contribution.}
    \label{fig:mass_contributions}
\end{figure*}

Figure \ref{fig:mass_contributions} decomposes the mass budget of the stars (top) and DM (bottom) in the $z\approx0$ cluster. Reading from left to right, the cluster material is divided into assembly channels, and then recombines to show how each assembly channel contributes to the $z\approx0$ satellites, BCG and ICL.

The assembly for both stars and DM are split into two broad channels: "early assembly" and "late assembly". Early assembly encompasses material which has already entered, or is in the process of entering, the cluster in the first snapshot at $z\approx2$. Conversely, late assembly encompasses material which enters the cluster from $z\approx2$ onwards. The following subsections define these channels and their sub-channels, and assess their mass fractions and contributions of their material to the $z\approx0$ components to check these are sensible and align with expectations of where their material should generally end up at $z\approx0$. We summarise our definitions in Table \ref{tab:definitions}.

\subsubsection{Early assembly}

We begin by defining the material in the early assembly channel. The first step is to identify structures which have either already merged or are in the process of merging with the cluster (haloes with their edge, $R_\textrm{178, halo}$, within the cluster $R_\textrm{178}$). These are designated as "original subhaloes". Any material within $R_\textrm{178}$ or part of these original subhaloes in the first snapshot is categorised as "in $z\approx2$ cluster".

Though DM linked to these early structures constitutes the entire early assembly channel for DM (aside from a small correction made at the end of the Section\,\ref{sec:late_assembly_dm}), additional star formation can occur within the cluster -- some of which we also link to early assembly. Stars can form in objects accreted before $z\approx2$ (termed "in-situ in original subhaloes", shortened to "in-situ orig. subhaloes" in Fig.~\ref{fig:mass_contributions}). Stars can also form outside of subhaloes (i.e. in either the BCG or ICL; termed "in-situ in cluster"). This is only a small ($\sim2$\,per cent) contribution, and the material almost entirely resides in the BCG at $z\approx0$, suggesting in-situ ICL formation is a negligible channel to ICL production (also found by \citealt{Brown_2024} in \textsc{Horizon-AGN}). We designate both these categories of in-situ star formation as sub-channels of early assembly. Any star formation in objects accreted after $z\approx2$ is instead linked to late assembly (see Section~\ref{sec:late_assembly_stars}). 

The two rightmost columns of Fig.~\ref{fig:z0_map} shows an example of the $z\approx0$ spatial distributions of the early and late assembly material in the BCG+ICL (top) and main halo (bottom). For both the stars and DM, the early assembly material is more centrally-concentrated than the late assembly material, reflecting the inside-out growth of haloes.

\subsubsection{Late assembly}\label{sec:late_assembly}

The material not assigned to the early assembly channel is attributed to the late assembly channel. The hierarchical assembly of cluster haloes means mergers with other haloes should be an important contribution to the stellar and DM content of the cluster, so we must first identify mergers.

\paragraph{Mergers:}\label{sec:late_assembly_mergers}

Merger identification is not straightforward, because substructure exists. To begin, we ignore the substructure status of haloes, and identify "candidate mergers" as events where an infalling halo's edge ($R_\textrm{178, halo}$) is first seen to cross the cluster $R_\textrm{178}$. In other words, the distance between the centres of the cluster and the infalling halo are greater than $R_\textrm{178, halo} + R_\textrm{178}$ in one snapshot (the "pre-merger snapshot") and then less than $R_\textrm{178, halo} + R_\textrm{178}$ in the subsequent snapshot ($\sim250$ Myr later in the "merger snapshot"). The merger time is given by that of the merger snapshot, the merger mass is the $M_\textrm{178, halo}$ in the pre-merger snapshot, and the merger mass ratio is ${M_\textrm{ratio}=M_\textrm{178}/M_\textrm{178, halo}}$ in the pre-merger snapshot. The merger mass and $M_\textrm{ratio}$ use the properties of the cluster and infalling haloes in the pre-merger snapshot to reduce the effect of mass loss via tidal stripping on the infalling halo's mass.\footnote{Ideally, we would use the maximum mass of the infalling halo prior to the merger, as this reflects the "true" mass of the halo before dynamic interactions occur and the \textsc{AdaptaHOP} structures begin to link. However, it is then ambiguous at which snapshot the cluster mass should be measured, and therefore also the merger time.}

Next, we apply the scheme to assign stellar and DM material to each candidate merger (see Sections \ref{sec:late_assembly_stars} and \ref{sec:late_assembly_dm} for details). We then check if each infalling halo is substructure of another halo (excluding the cluster halo itself) in the pre-merger snapshot. If an infalling halo is not substructure, then we call this a merger. Conversely, if an infalling halo is substructure (i.e. it has a "host" halo in the pre-merger snapshot), then it is instead grouped together with the merger of its host halo. This means that e.g. an infalling galaxy group will be counted as a single merger with the merger properties of the group halo, encompassing all the material attributed to both the group halo and the subhaloes.

\begin{figure*}
	\includegraphics[width=\textwidth]{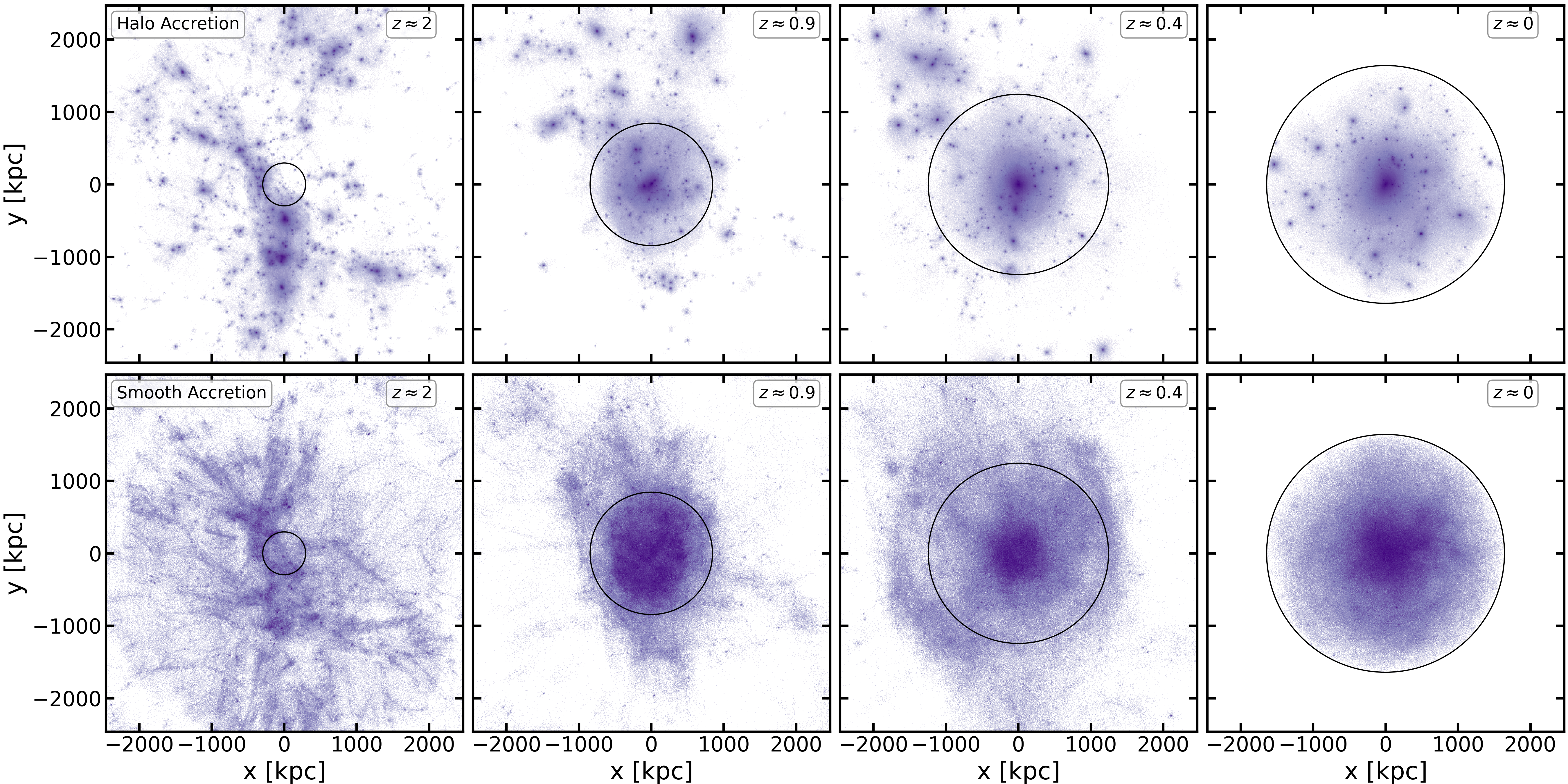}
    \caption{Projected surface density map of the DM in the late assembly channel at intervals of decreasing redshift (left to right) along two arbitrary spatial axes centred on the main progenitor halo for Hrz049. The top and bottom rows contain the DM material in the two sub-channels of late assembly, the halo and smooth accretion respectively. In all panels, darker colours represent higher projected surface densities on a logarithmic scale. The black circles indicate $R_{178}$ around the centre.}
    \label{fig:redshifts_map}
\end{figure*}

\paragraph{Stars:}\label{sec:late_assembly_stars}

Using the structures defined previously in Section \ref{sec:merger_trees}, we find almost all star particles in our simulation cutouts are assigned to haloes. This is unsurprising, given that significant star formation only occurs in haloes above the \textsc{AdaptaHOP} resolution limit, and that the stellar component is typically more centrally-concentrated than the DM component due to physics of gas collapse. Aside from the aforementioned in-situ star formation, we therefore also expect stars to only assemble into the cluster through halo mergers, so the sole sub-channel of late assembly for the stellar material is "halo accretion" (and therefore these terms can be used interchangeably for the stellar material).

Within the halo accretion sub-channel, the stellar material which enters through identified mergers is assigned to the "direct accretion" sub-channel. Star particles are attributed to candidate mergers by being part of the infalling halo in both the pre-merger and merger snapshot, with assignment priority is given to the earliest candidate mergers if necessary.

Star formation may still occur in these merged haloes after cluster infall. This stellar material constitutes the final in-situ sub-channel, "in-situ in accreted subhaloes" (shortened to "in-situ accr. subhaloes in Fig.~\ref{fig:mass_contributions}) as part of the halo accretion sub-channel, and is defined as any star particle that forms within $R_{178}$ inside a subhalo that is not one of the original subhaloes. We do not attempt to attribute this to specific mergers, as we do not track gas particles nor post-infall haloes. This is because tracking substructure within haloes is difficult when there is a relatively small density contrast between the substructure and the background material in the halo, making merger tree branches more susceptible to breaking, especially in the dense inner regions of the cluster.

Combined, the direct accretion and "in-situ in accreted subhaloes" sub-channels encompass almost all of the stellar material associated with the halo accretion sub-channel, demonstrating that our merger identification scheme is successful. The negligible amount ($<1$ per cent) of stellar material that is not assigned to either (from e.g. poorly-resolved low-mass satellites) is put into the "unassigned" sub-channel.

\paragraph{DM:}\label{sec:late_assembly_dm}

The halo accretion sub-channel is also present for the DM, and DM material is assigned to this sub-channel in an identical way to that described for the stellar material in Section \ref{sec:late_assembly_stars}. Unlike the stellar material however, much of the DM material genuinely exists outside of collapsed structures or in unresolved haloes, and DM particles on bound orbits of haloes extends spatially beyond the \textsc{AdaptaHOP} structure definitions. This material assigned to halo accretion sub-channel is therefore a relatively conservative attribution, so we also estimate if the DM material surrounding each candidate merger is bound to the infalling halo.

We compute the potential generated by the DM within $2\,R_\textrm{178, halo}$ of each infalling halo's centre in the pre-merger snapshot for the candidate mergers, following the \cite{Barnes_Hut_1986} method\footnote{Our implementation can be found at: \url{https://github.com/garrethmartin/bh_potential}.}. We then calculate the binding energy of each DM particle within $3\,R_\textrm{178, halo}$, and attribute the bound material to that candidate merger. Assignment priority is given to the candidate merger that each particle is most tightly bound to, within a 1\,Gyr window of the earliest-occurring merger that each particle is considered for.

The difficulty in defining the extent of infalling haloes in cosmological simulations is a reflection of the structures often themselves being in the process of collapse, and being embedded in wider cosmic structure, making any outer boundary somewhat arbitrary. There is therefore no "correct" choice for how to define the boundness/attribution of material to a halo.

The remaining DM material is assigned to the "smooth accretion" sub-channel. This is a substantial fraction of the cluster content ($\sim30$ per cent), in line with findings from other cosmological simulations \citep[e.g.][]{Genel_2010}. Figure \ref{fig:redshifts_map} shows the halo (top) and smooth (bottom) accretion for a single cluster at a series of snapshots from $z\approx2$ to $z\approx0$. At all redshifts, the smooth accretion is clearly a more diffuse component, and can be interpreted as the cosmic web outside of collapsed structures. We note this smooth accretion sub-channel does not simply arise from apparent cluster growth through pseudo-evolution, nor the contribution of unresolved haloes due to limited simulation resolution (see Sections\,\ref{sec:discussion_assembly_time} and \ref{sec:discussion_smooth_accretion} respectively).

Visual inspection reveals a sharp influx of smooth accretion material entering the cluster in the first few snapshots. These are DM particles on high energy orbits that happen to be outside of $R_\textrm{178}$ in the first snapshot, rather than a genuine spike in smooth accretion at $z\approx2$. We therefore perform a small correction to move the smooth accretion material that enters the cluster $R_\textrm{178}$ in the first three snapshots ($<2$ per cent of the total DM material) to the "in $z\approx2$ cluster" sub-channel.

\subsection{Defining the orbital properties}\label{sec:orbital_properties}

The orbital properties of particles can be fully described by their orbital energy and orbital angular momenta. We use the specific versions of these quantities (dividing by the particle mass) to allow comparison between the stars and DM particles, since a typical DM particle will have a higher orbital energy or angular momentum than a typical star particle simply due to its mass ($M_\textrm{DM}/ M_\textrm{star}\,\approx\,40$). We calculate the specific total energy ($\varepsilon$) as the sum of the specific kinetic energy ($\varepsilon_\textrm{k}$) and specific potential energy ($\varepsilon_\textrm{p}$), defined as
\begin{equation}
   \varepsilon = \varepsilon_\textrm{k} + \varepsilon_\textrm{p} = \frac{v^2}{2} + (\phi - \phi_\textrm{min}),
    \label{eq:specific_energy}
\end{equation}

where $v$ is the particle speed, $\phi$ is the potential at the position of the particle, and $\phi_\textrm{min}$ is the potential at the position of the most bound particle in the cluster. The normalisation to $\phi_\textrm{min}$ means the specific energy will always be positive, thus the usual implication that a positive total energy corresponds to an unbound particle does not apply. In a static potential, $\varepsilon$ would be constant throughout the orbit for a given particle. The specific angular momentum is given by
\begin{equation}
   j = \left| \bf{r} \times \bf{v} \right|,
    \label{eq:specific_angular_momentum}
\end{equation}
where $\bf{r}$ and $\bf{v}$ are the particle position and speed respectively relative to the bulk position/velocity of the cluster. Unlike the instantaneous position and velocity of a given particle, which changes rapidly over the course of an orbit, the orbital energy and angular momentum remain approximately constant throughout the orbit (in the absence of dynamical evolution), making these suitable parameters. This property also naturally helps establish a connection between an individual particle's dynamics at cluster infall and at $z\approx0$.

The mean specific energy ratio $\langle \varepsilon_\textrm{ratio} \rangle$ and specific angular momentum ratio $\langle j_\textrm{ratio} \rangle$ between the BCG+ICL stars and main halo DM are key quantities that have been shown to be less than unity in \citetalias{Butler_2025}. A key aim of this paper is to gain physical insights into the origin of these discrepancies, by decomposing the $z\approx0$ material into assembly channels and further into individual mergers. The following formulae describe $\varepsilon$, but the equivalent formulae can be written for $j$. We express the net $\langle \varepsilon \rangle$ (e.g. for stars) as
\begin{equation}
    \langle \varepsilon \rangle = \sum_i f_i \langle \varepsilon_i \rangle,
\end{equation}
where $\langle \varepsilon_i \rangle$ is the mean specific energy of the material in assembly channel $i$, and the mass fraction $f_i$ in that assembly channel acts as a weight. $\langle \varepsilon_\textrm{ratio} \rangle$ can therefore be expressed as
\begin{equation}
    \langle \varepsilon_\textrm{ratio} \rangle= \frac{\langle \varepsilon_{\rm stars} \rangle}{\langle \varepsilon_{\rm DM} \rangle} = \frac{\sum_i f_{\textrm{stars},i} \langle \varepsilon_{\rm stars,i} \rangle}{\sum_j f_{\textrm{DM},j} \langle \varepsilon_{\rm DM,j} \rangle}.
\end{equation}
The general method of decomposing $\langle \varepsilon \rangle$ in this manner is applicable for any arbitrary segregation of the material. However, the $f_c$ and $\langle \varepsilon_c \rangle$ of a given assembly channel $c$ can give insight into the physical mechanisms that are relevant for the associated material.\footnote{We use $c$ here to denote an arbitrary assembly channel, but since the stars and DM are not necessarily split into the same set of assembly channels, $i$ and $j$ are used respectively in the formulae.} The mass fraction of a given channel indicates how important that channel is, and comparing the mass fractions of the same channel in DM and stars indicates if that channel is proportionally more important to either. Comparing $\langle \varepsilon \rangle$ between different channels of e.g. stars demonstrates how/if the material in these channels are distributed differently, and taking ratios between $\langle \varepsilon \rangle$ of the same channel in stars and DM gives hints on how physical mechanisms may affect stars and DM differently. Finally, the effect of masking the BCG on all these quantities must be considered to understand how this influences the net $\langle \varepsilon_\textrm{ratio} \rangle$.

\section{Results}\label{sec:results}

\subsection{Assembly channels}\label{sec:results_assembly_channels}

To understand the bulk differences between the stars and DM at $z\approx0$, the orbital properties of the material from each assembly channel must be explored.

\begin{figure*}
	\includegraphics[width=\textwidth]{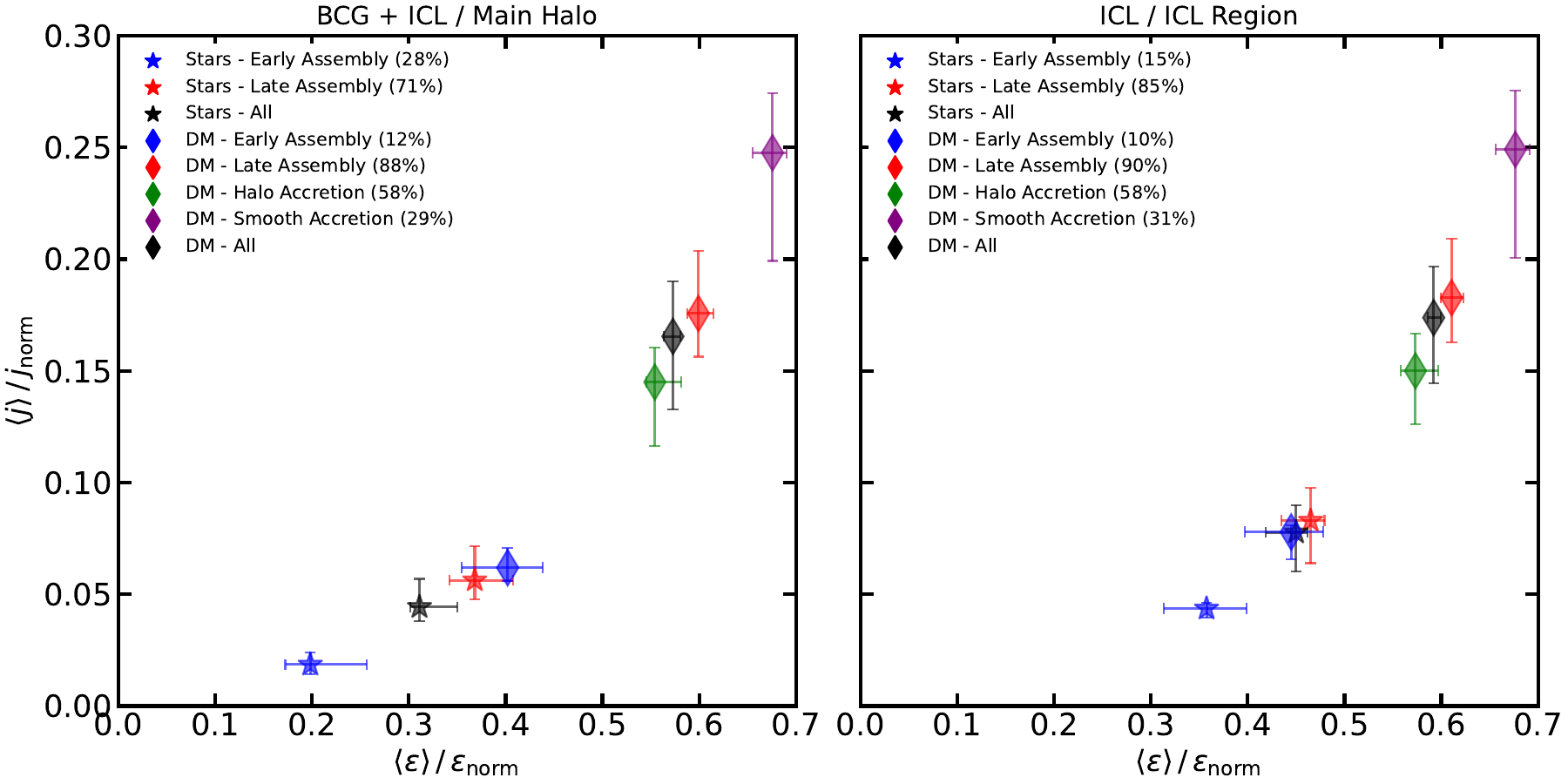}
    \caption{Mean specific angular momentum $\langle j \rangle$ against mean specific energy $\langle \varepsilon \rangle$ of each assembly channel (colours), for the BCG+ICL (left) and the ICL material (right). Stars and DM are represented by stars and diamonds respectively. $\langle \varepsilon \rangle$ and $\langle j \rangle$ are normalised for each cluster by the specific energy $\varepsilon_\textrm{norm}$ and specific angular momentum $j_\textrm{norm}$ of a test particle on a circular orbit at $R_{178}$. Each point represents the median between clusters, and the error bars the 16th to 84th percentile. The legend quotes the mean percentage that each assembly channel contributes to the mass budget.}
    \label{fig:se_am_assembly_channel}
\end{figure*}

\begin{figure*}
	\includegraphics[width=\textwidth]{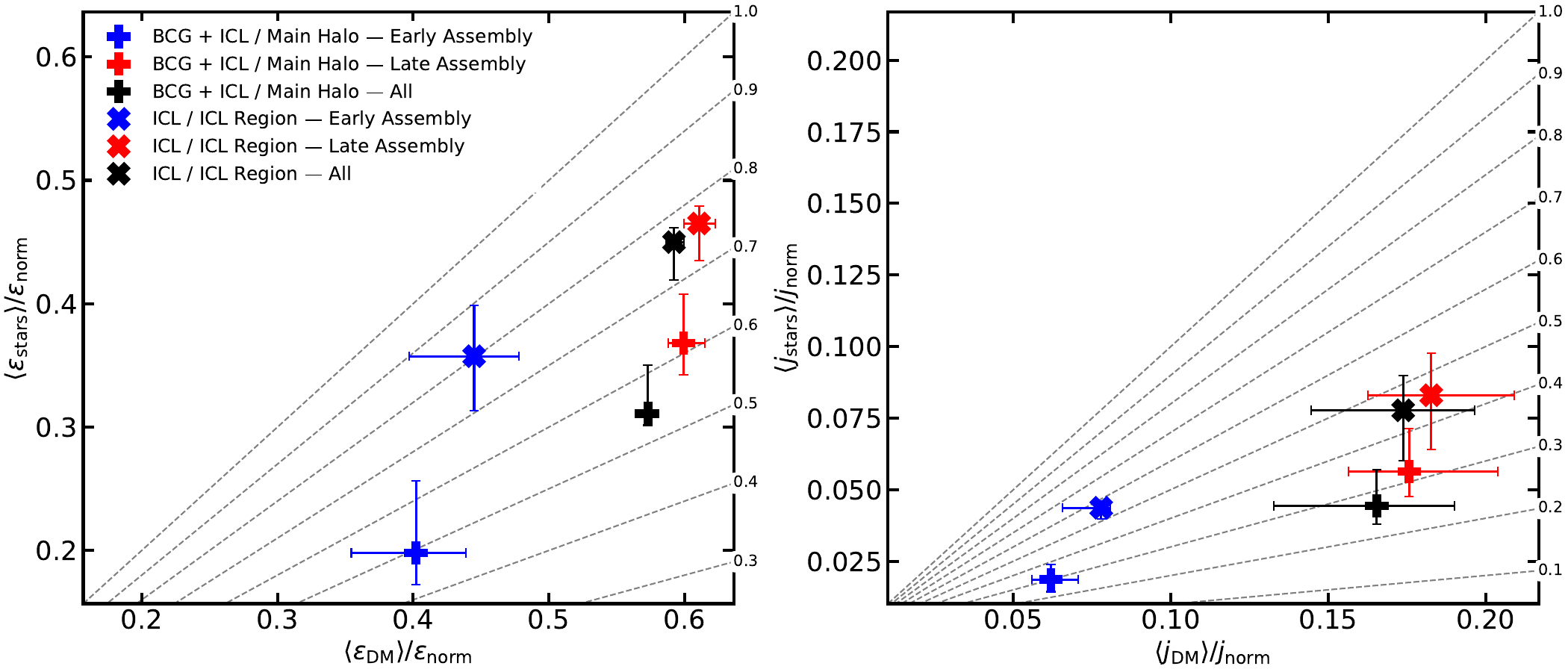}
    \caption{Mean specific energy of stars $\langle \varepsilon \rangle$ against DM (left) and mean specific angular momentum $\langle j \rangle$ of stars against DM (right) for each assembly channel. The BCG+ICL and ICL material are represented by pluses and crosses respectively. Each point represents the median between clusters, and the error bars the 16th to 84th percentile. The dashed grey lines guide the eye to the ratio between stars and DM.}
    \label{fig:se_am_stars_vs_dm}
\end{figure*}

Figure \ref{fig:se_am_assembly_channel} shows the mean specific orbital angular momentum $\langle j \rangle$ against specific orbital energy $\langle \varepsilon \rangle$ at $z\approx0$ of the BCG+ICL (left) and  ICL (right), split into assembly channels. For each cluster, $\langle \varepsilon \rangle$ and $\langle j \rangle$ are normalised by the specific orbital energy and angular momentum of a test particle on a circular orbit at the cluster $R_{178}$, which we label as $\varepsilon_\textrm{norm}$ and $j_\textrm{norm}$ respectively. The displayed points and error bars are the median and $16^\textrm{th}$ to $84^\textrm{th}$ percentiles between clusters. The legend also gives the mean percentage contribution from each assembly channel, as seen in Fig.~\ref{fig:mass_contributions}.

We focus first on the ICL (right subplot). The $\langle \varepsilon \rangle$ and $\langle j \rangle$ of the stellar material from all assembly channels (black star) are lower than the DM material from all assembly channels (black diamond). This is also true for both the early (blue) and late (red) assembly material separately. Next, we see for both stars and DM that the early assembly material occupies significantly lower $\langle \varepsilon \rangle$ and $\langle j \rangle$ orbits than the late assembly material. When the DM late assembly material is further divided into smooth (purple diamond) and halo (green diamond) accretion, we see that the smooth accretion material is typically on the highest $\langle \varepsilon \rangle$ and $\langle j \rangle$ orbits. This implies the DM material entering through the smooth accretion sub-channel typically loses little, if any, energy or angular momentum after entering the cluster.

If the BCG region is not masked, such that the BCG+ICL is instead considered (left subplot), the stellar material from both the early and late assembly channels occupy even lower $\langle \varepsilon \rangle$ and $\langle j \rangle$. This is because the central regions of the cluster are biased towards encompassing particles on lower energy and angular momentum orbits, which are preferentially removed when the BCG is masked. This effect is strongest for the stars in the early assembly channel, since the majority of these stars end up in the BCG. The same is true for the DM material but to a much lesser degree, due to the significantly smaller contribution of the material in the BCG region to the main halo mass budget. This means the $\langle \varepsilon \rangle$ and $\langle j \rangle$ of the DM material is relatively insensitive to the masking of the BCG region.

Figure \ref{fig:se_am_stars_vs_dm} is complimentary to Fig.~\ref{fig:se_am_assembly_channel}, plotting the $\langle \varepsilon \rangle$ of the stellar material against the DM material in each assembly channel on the left panel (normalised by $\varepsilon_\textrm{norm}$). The BCG+ICL and ICL are represented by pluses and crosses respectively. The equivalent for angular momentum is plotted on the right panel. We do not further divide late assembly into halo and smooth accretion, since smooth accretion does not have an equivalent in the stellar component. The dashed grey lines trace constant values of the star-to-DM ratio, in steps of 0.1. These provide a visual guide to the ratio between the $\langle \varepsilon \rangle$ or $\langle j \rangle$ of the stellar and DM material from each channel.

Regardless of the BCG masking or assembly channel, the stars always have lower $\langle \varepsilon \rangle$ and $\langle j \rangle$ than the DM, such that all points lie below the 1:1 dashed line. This indicates that the processes responsible for the $\langle \varepsilon \rangle$ and $\langle j \rangle$ of the stars being lower than the DM affect both the early and late assembly material, and that both the early and late assembly channels are contributing to the overall $\langle \varepsilon \rangle$ and $\langle j \rangle$ ratios being lower than unity. For a given assembly channel, we again see the BCG region masking does not greatly affect the mean orbital properties of the DM, but it does impact the stars, increasing when the BCG is masked. We see the main results from \citetalias{Butler_2025} reproduced here, that $\langle \varepsilon_\textrm{ratio} \rangle\approx0.75$ for the ICL and $\langle \varepsilon_\textrm{ratio} \rangle\approx0.55$ for the BCG+ICL.

One interpretation of these results is that the stratification in energy and angular momentum is a reflection of the initial boundness of the material to infalling haloes. The stellar material is initially the most tightly-bound to infalling haloes, typically occupying their central regions, followed by the DM material in the halo accretion sub-channel, and finally the DM material in the smooth accretion sub-channel is either very loosely bound or unbound to any collapsed structure upon cluster infall. This is the order these components occupy in both panels of Fig.~\ref{fig:se_am_assembly_channel}, from lowest to highest $\langle \varepsilon \rangle$ and $\langle j \rangle$. The two are connected through tidal stripping, which removes material from an infalling halo roughly in order of how weakly bound it is. How tightly a particle is bound to its infalling halo determines when it is stripped, and therefore how much orbital decay, if any, the halo has undergone by the time that material is released into the cluster. We discuss this further in Section \ref{sec:discussion}.

To explore this hypothesis in more detail, we investigate individual mergers, which contribute material to the halo accretion sub-channel. This tests if this differential stripping is indeed occurring on a per-merger basis, and if there are any dependencies of this process on merger properties.

\subsection{Halo accretion}\label{sec:results_halo_accretion}

\begin{figure*}
	\includegraphics[width=\textwidth]{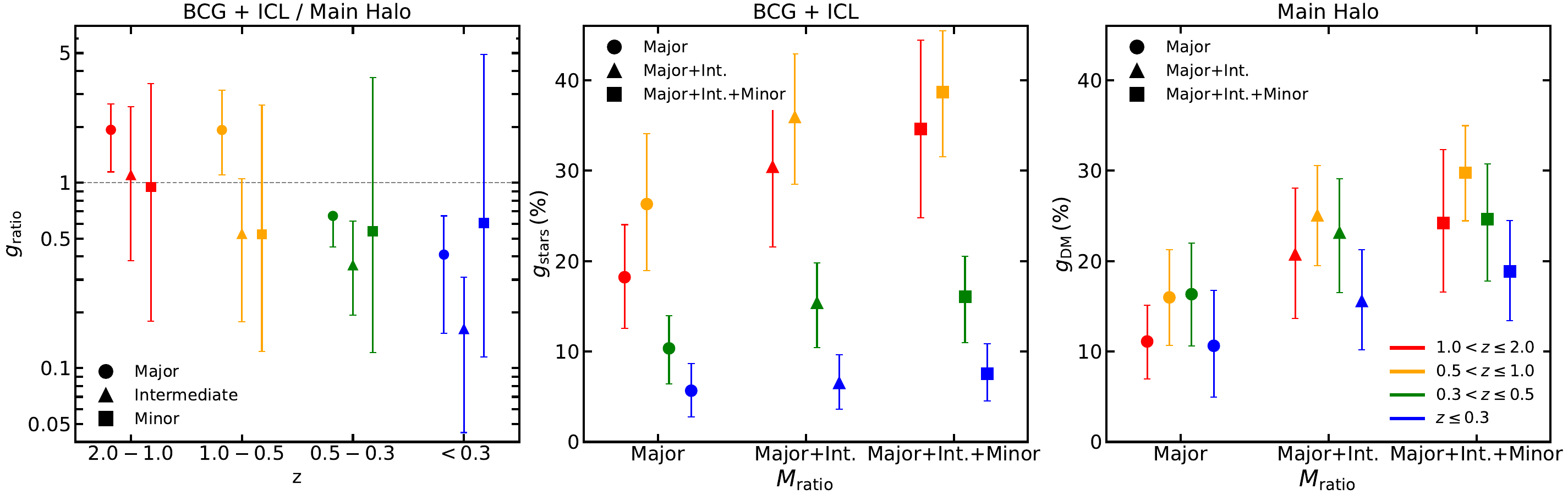}
    \caption{\textit{Left}: Ratio of the percentage mass contribution to the BCG+ICL and the percentage mass contribution to the main halo, $g_\textrm{ratio}$, from each merger. Mergers are binned by merger mass ratio $M_\textrm{ratio}$ (shapes) and merger redshift $z$ (colours/x-axis position). The points give the median and the error bars give the 16th to 84th percentile. \textit{Middle}: The total stellar mass contribution to the BCG+ICL of all mergers in a given merger bin, $\Sigma g_\textrm{stars}$. The mean between clusters and the bootstrapped error on the mean are shown as the points and error bars respectively. \textit{Right}: The equivalent of the middle panel for the total DM mass contribution to the main halo of all mergers in a given merger bin, $\Sigma g_\textrm{DM}$.}
    \label{fig:merger_contributions_bcg_icl}
\end{figure*}

Figures \ref{fig:se_per_merger_bcg_icl} and \ref{fig:merger_contributions_bcg_icl} explore the halo accretion sub-channel for the BCG+ICL material on a per-merger basis. We focus on orbital energy here, but note that these findings are qualitatively reproduced for orbital angular momentum. The results shown below are for the combined BCG+ICL, as merger-sourced material can be strongly affected by orbital decay. We provide versions of these figures limited to exclusively the ICL in Appendix \ref{sec:appendix_additional_results}.

\subsubsection{Per-merger energy ratios}\label{sec:results_energy_ratios}

\begin{figure}
	\includegraphics[width=\columnwidth]{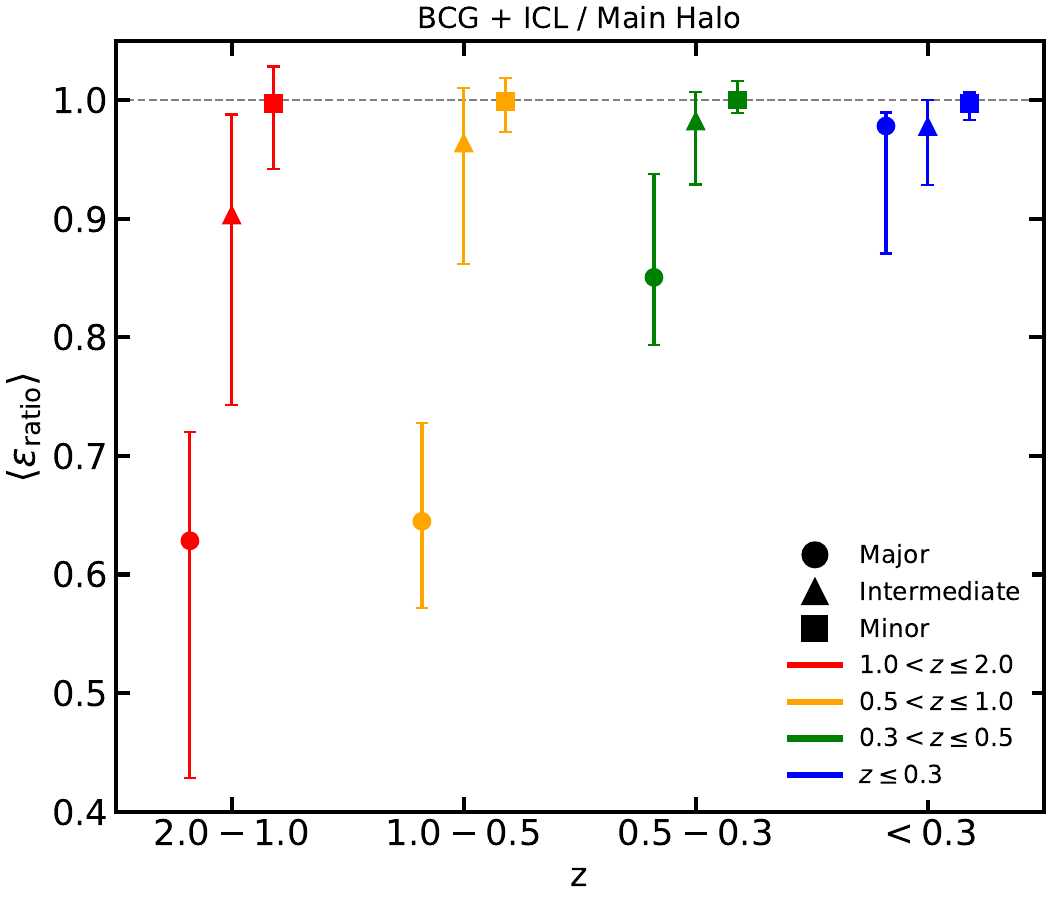}
    \caption{Mean specific energy ratio at $z\approx0$ between the BCG+ICL and main halo material, $\langle \varepsilon_\textrm{ratio} \rangle$, from each merger. Mergers are binned by merger mass ratio $M_\textrm{ratio}$ (shapes) and merger redshift $z$ (colours/x-axis position). The points give the median and the error bars give the 16th to 84th percentile.}
    \label{fig:se_per_merger_bcg_icl}
\end{figure}

In Fig.~\ref{fig:se_per_merger_bcg_icl} we take the $\langle \varepsilon_\textrm{ratio} \rangle$ between the BCG+ICL stars and DM from each merger. Mergers are binned by both mass ratio $M_\textrm{ratio}$ and redshift $z$, and for each bin we plot the median $\langle \varepsilon_\textrm{ratio} \rangle$ along with the 16th to 84th percentiles. We classify mergers as major ($M_\textrm{ratio}\leq10$), intermediate ($10<M_\textrm{ratio}\leq100$) or minor ($M_\textrm{ratio}>100$). The four merger redshift bins ($1.0<z\leq2.0$, $0.5<z\leq1.0$, $0.3<z\leq0.5$ and $z\leq0.3$) each encompass approximately 2.4\,Gyr. We note that the methodology used to define the merger properties will systematically overestimate $M_\textrm{ratio}$ (see Section \ref{sec:late_assembly_mergers}); the bin edges have a higher value than is typical in literature to account for this.

We see that $\langle \varepsilon_\textrm{ratio} \rangle$ depends strongly on $M_\textrm{ratio}$. The medians for all of the minor mergers are $\langle \varepsilon_\textrm{ratio} \rangle \approx1$, showing no systematic offset between $\langle \varepsilon_\textrm{DM} \rangle$ and $\langle \varepsilon_\textrm{stars} \rangle$. We also see that medians for the intermediate mergers lie between 0.9 and 1, albeit often with a large dispersion enveloping $\langle \varepsilon_\textrm{ratio} \rangle = 1$, but also reaching down to $\langle \varepsilon_\textrm{ratio} \rangle <0.75$ for the earliest mergers. This suggests there is some degree of differential stripping of stars and DM occurring for these haloes, which is increasingly realised for mergers at earlier times. Finally, we see that major mergers typically have the lowest $\langle \varepsilon_\textrm{ratio} \rangle$ in each $z$ bin. There is a strong dependence of $\langle \varepsilon_\textrm{ratio} \rangle$ on $z$, reaching a median of $\langle \varepsilon_\textrm{ratio} \rangle \approx0.65$ for the two highest $z$ bins. Together, these trends point to the presence of differential stripping of stars and DM in more massive infalling haloes, an effect that tends to be stronger for mergers that occurred earlier.

However, in Fig.~\ref{fig:se_am_stars_vs_dm} the late assembly BCG+ICL has $\langle \varepsilon_\textrm{ratio} \rangle \approx 0.6$ (red plus), which is below the $\langle \varepsilon_\textrm{ratio} \rangle$ for almost all individual mergers. Differential stripping within mergers therefore cannot account for the full phase-space offset on its own, reinforcing that the additional component of smooth accretion is necessary to fully account for the orbital differences between the stars and DM.\footnote{Another explanation would be that the material from the major mergers at $1.0<z\leq2.0$ is completely dominant, but we will shortly see this is not the case.}

Appendix \ref{sec:appendix_additional_results} shows the same information for the ICL. The most notable change is a substantial increase in $\langle \varepsilon_\textrm{ratio} \rangle$ for early major mergers, with the medians rising to $\langle \varepsilon_\textrm{ratio} \rangle \approx 0.8$ and $\langle \varepsilon_\textrm{ratio} \rangle \approx 0.75$ for the $1.0<z\leq2.0$ and $0.5<z\leq1.0$ bins respectively. This can be attributed to the stellar material on low energy orbits being preferentially located within the BCG, which is then subsequently masked and thus does not contribute to $\langle \varepsilon_\textrm{ratio} \rangle$. The strongest impact is on mergers which contribute a significant fraction of their stellar material to the BCG, which is the case for early major mergers. All of these trends are also qualitatively reproduced for the angular momentum ratios (not shown).

These results demonstrate that the $M_\textrm{ratio}$ and merger $z$ are important for the degree of differential stripping of stars and DM from satellites, being strongest for earlier and closer mass ratio mergers. This leads to stars being deposited at lower $\langle \varepsilon \rangle$ than the DM originating from the same merger, contributing to $\langle \varepsilon_\textrm{ratio} \rangle <1$ for the halo accretion sub-channel.

\subsubsection{Mass contributions}\label{sec:results_mass_contributions}

Although the $\langle \varepsilon_\textrm{ratio} \rangle$ for each merger is informative about the dynamical evolution and differential stripping, this needs to be combined with the mass contribution of each merger to the BCG+ICL to understand how important those mergers are to the $\langle \varepsilon_\textrm{ratio} \rangle$ of the halo accretion channel.

We define $g$ as the percentage mass contribution of a given merger to the $z\approx0$ component of interest, relative to the total merger contribution to that same component of its host cluster. For DM, the total merger contribution is the halo accretion material, but for stars it is just the direct accretion material, since the "in-situ in accreted subhaloes" material is not attributed to specific mergers. $g_\textrm{ratio} = g_\textrm{stars} / g_\textrm{DM}$ is analogous to $\langle \varepsilon_\textrm{ratio} \rangle$, being the relative mass contributions of the stars to DM on a per-merger basis. We define $\Sigma g$ as the percentage sum of all mergers in a given bin to the component of interest.

The left panel of Fig.~\ref{fig:merger_contributions_bcg_icl} shows the $g_\textrm{ratio}$ of each merger bin for the BCG+ICL, with the median and 16th to 84th percentiles displayed as points and error bars respectively. Most mergers lie below the $g_\textrm{ratio}=1$ dashed line, meaning they contribute a greater relative amount of DM material to the main halo than stellar material to the BCG+ICL. On the other hand, the major mergers in the two highest $z$ bins, and approximately half of the intermediate mergers in the $1.0<z\leq2.0$ bin, contribute more significantly to the BCG+ICL than to the main halo. 

The middle and right panel of Fig.~\ref{fig:merger_contributions_bcg_icl} show $\Sigma g_\textrm{stars}$ to the BCG+ICL and $\Sigma g_\textrm{DM}$ to the main halo respectively against $M_\textrm{ratio}$. Circles represent the major mergers, triangles represent major plus intermediate mergers, and squares represent all mergers in the $z$ bin. There is a large scatter between clusters for the $\Sigma g$ of each merger bin due to different assembly histories, in particular for major mergers, since some clusters have several major mergers in a given $z$ bin and others have none.\footnote{Additional factors which drive genuine scatter of $g$ between mergers in the same bin include: different (and evolving) ranges of infalling halo masses due to differences in cluster progenitor masses; the progression of stellar and DM material liberation from the infalling halo; and for $g_\textrm{stars}$, scatter in the stellar-to-halo mass relation.} We therefore plot the mean between clusters and the bootstrapped error on the mean as the points and error bars respectively.

In all $z$ bins for both the stars and DM, we observe that the majority mass contribution comes from the major mergers, with $\Sigma g$ plateauing as progressively less-equal merger ratios are included. Since DM is stripped from infalling haloes before their stars, a significant fraction of the accreted DM can already have been incorporated into the main halo while the associated stars remain bound to their satellite galaxies and have yet to contribute to the BCG+ICL. This distinction is not simply driven by differences in the stellar-to-halo mass ratios of galaxies across merger mass ratios: when bound satellite stars are included (i.e. the satellites are not masked), $g_\textrm{stars}$ closely follows $g_\textrm{DM}$ across the merger bins. The divergence therefore arises from the differential stripping of the two components, rather than from differences in their initial relative abundances. The material from the two highest $z$ bins collectively dominate the BCG+ICL (direct accretion) mass budget, driven by the major and intermediate mergers, whilst the main halo (halo accretion) mass budget is more evenly split between $z$ bins.

\begin{figure*}
	\includegraphics[width=\textwidth]{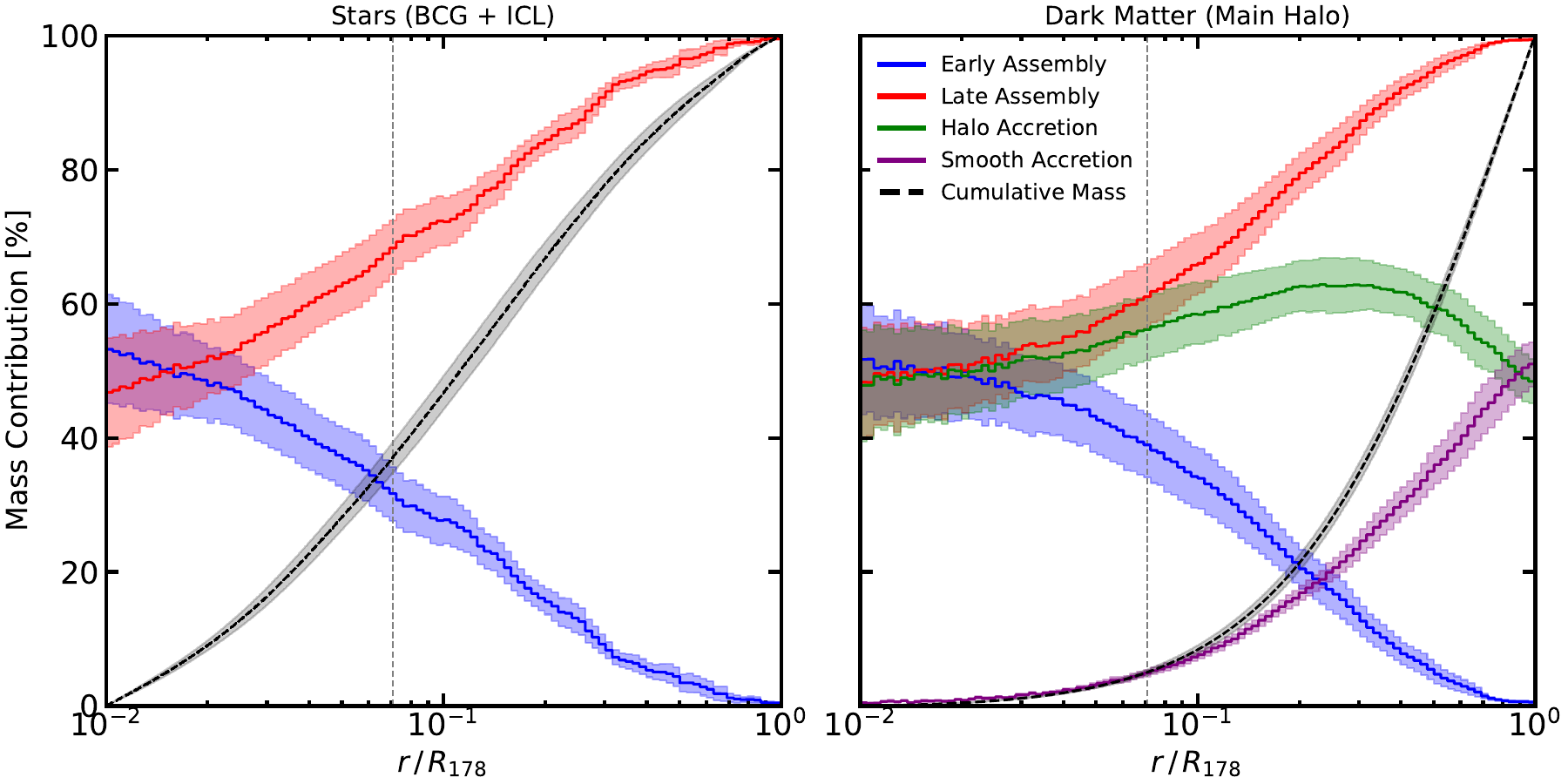}
    \caption{Percentage mass contribution of assembly channels to the $z=0$ BCG+ICL stars (left) and main halo DM (right) against cluster-centric radius, normalised by $R_{178}$. The stars and DM are both divided into early and late assembly, with the late assembly for the DM also split into halo and smooth accretion. Additionally, the dashed black curve displays the cumulative mass with increasing radius, normalised to the mass within $R_{178}$. All curves shown represent the median between clusters, and the shaded regions represent the bootstrapped error on the mean. The dashed vertical grey line represents the mean 100\,kpc boundary between the BCG and ICL.}
    \label{fig:radial_contributions}
\end{figure*}

Despite typically containing the highest infalling halo mass mergers, the lowest $z$ bin has the lowest mass contributions for both stars and DM. This is partly due to material being located beyond the cluster $R_{178}$, either not having yet entered (due to the merger time being defined when the infalling halo $R_{178}$ crosses the cluster $R_{178}$, which for major mergers in particular can be before most material has entered) or having exited the cluster $R_{178}$ after reaching first pericentre ("backsplash"). Again for the stellar material, very little or any tidal stripping has occurred, thus the intermediate and minor mergers contribute almost no stars.

Masking the BCG (Appendix\,\ref{sec:appendix_additional_results}) substantially decreases the stellar mass contribution from early major mergers, both on a per-merger basis and for their total contribution to the BCG+ICL direct accretion. This incorporation of stellar material from early major mergers into the BCG is well-established \citep[e.g.][]{Brown_2026, Pearce_2026}. Since $g_\textrm{stars}$ is a percentage rather than absolute mass contribution, the relative importance of all other merger bins rise. The DM mass contributions are largely insensitive to the masking of the BCG region.

Taken together, these results demonstrate a clear distinction between the assembly of the DM and stellar components of clusters. While mergers across all epochs and mass ratios are able to contribute significant DM to the main halo, intermediate and major mergers that occurred early enough ($z>0.5$) to have undergone substantial stellar stripping contribute much more significantly to the build-up of the BCG+ICL material. Since these merging haloes also deposit their stars at substantially lower energies than their DM material ($\langle \varepsilon_\textrm{ratio} \rangle <1$), the differential stripping mechanism is therefore responsible for much of the phase-space offset between the (BCG+)ICL stars and DM.

\subsection{Spatial distributions}\label{sec:results_spatial_distributions}

We have shown that there are substantial differences in the bulk orbital properties of the BCG+ICL and main halo material, both in total and across different assembly channels. We now investigate if (and how) these differences manifest in more directly-observable quantities, namely the radial density profiles and morphological shapes of BCG+ICL and main halo material, to understand how the material from different assembly channels influences the overall spatial distribution.

\subsubsection{Radial profiles}\label{sec:results_radial_profiles}

Figure \ref{fig:radial_contributions} displays the relative mass contribution of each assembly channel to the BCG+ICL (left) and main halo (right) material as a function of cluster-centric radius, with the curves representing the mean between clusters. For both the BCG+ICL and main halo, the early assembly material contribution is most important in the innermost regions of the cluster, accounting for approximately 50 per cent of the total material, becoming negligible beyond $0.7R_{178}$. The late assembly material therefore becomes increasingly important at larger radii. The clear stratification between the early and late assembly material by $z\approx0$ demonstrates the inside-out growth of the cluster, where later-accreted material is typically still found at larger radii. In other words, the memory of accretion time is not completely erased by cluster relaxation processes.

\begin{figure*}
	\includegraphics[width=\textwidth]{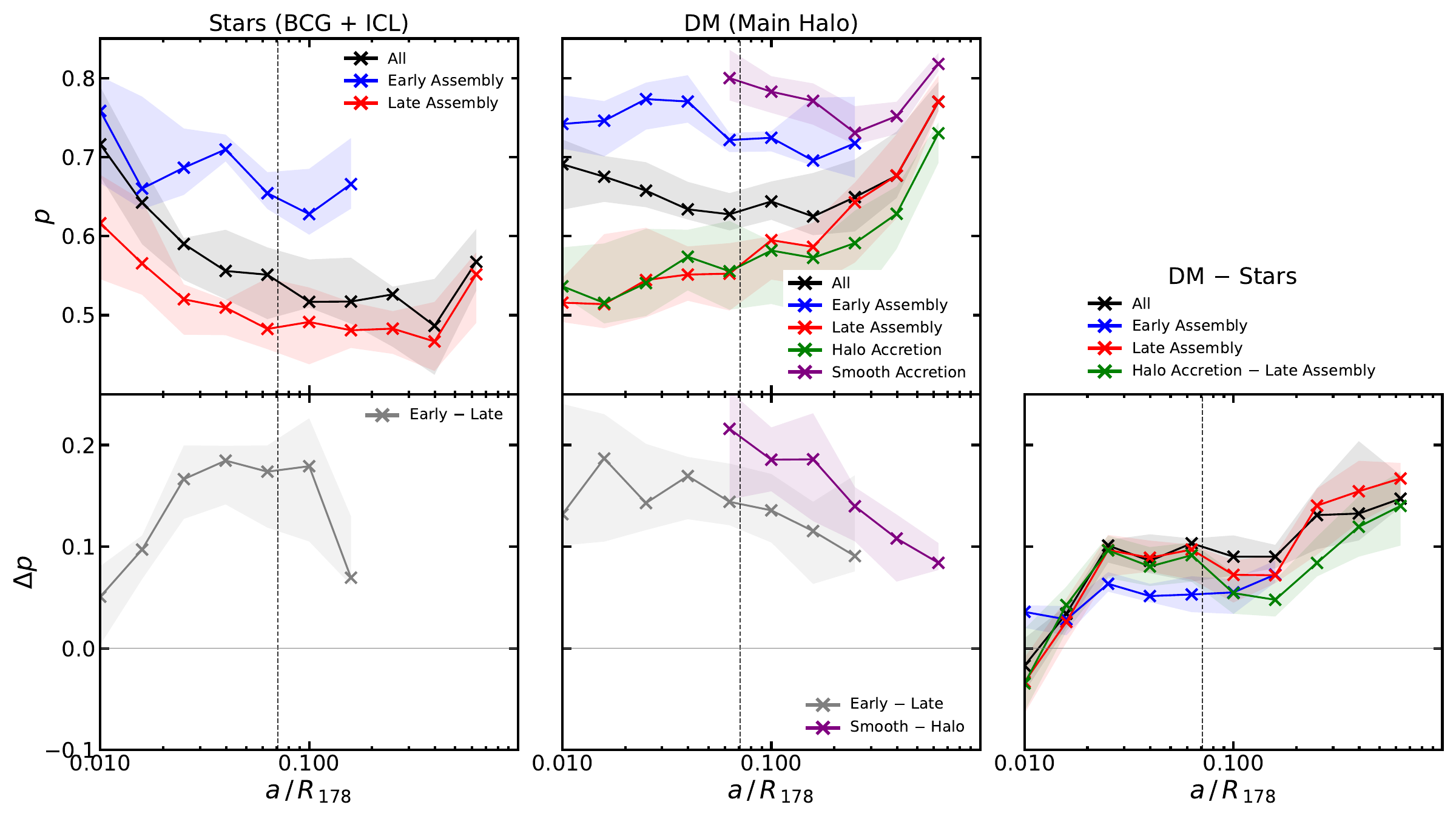}
    \caption{\textit{Top}: Morphological shape profiles of the BCG+ICL (left) and main halo DM (middle), characterised by the axis ratio $p$ as a function of the major axis length $a$, normalised by $R_{178}$. The points and connecting lines show the mean between clusters/projections and the bootstrapped error on the mean respectively. The dashed vertical grey line represents the mean 100\,kpc boundary between the BCG and ICL. \textit{Bottom}: Axis ratio difference $\Delta p$ between components on a per-cluster basis. The left panel shows $\Delta p$ between the BCG+ICL assembly channel shapes, the middle panel shows $\Delta p$ between the main halo assembly (sub-)channel shapes, and the right panel shows $\Delta p$ between the same assembly channels between the BCG+ICL and main halo.} 
    \label{fig:shapes}
\end{figure*}

The dashed black curves in Fig.~\ref{fig:radial_contributions} display the cumulative mass with increasing radius for the BCG+ICL (left) and main halo material (right). Approximately half of the BCG+ICL stellar material is within $0.1R_{178}$, but for the main halo DM material this percentage is $<10$ per cent. Additionally, we see that the BCG+ICL curve is beginning to plateau in the outer cluster regions, whereas the main halo has yet to show signs of plateauing. Given the logarithmic scaling of the radius, and that the cluster volume enclosed is proportional to the radius cubed, the plateau in the BCG+ICL cumulative mass implies there is very little stellar mass beyond $R_{178}$. In contrast, it is clear there is a significant amount of DM outside of $R_{178}$.

For the DM, the relative contributions of the two sub-channels of the late assembly channel, halo and smooth accretion, are also displayed. The halo accretion material is dominant over the smooth accretion material for the cluster inner regions, with the smooth accretion material contributing around 5 per cent of the main halo DM at $0.1R_{178}$ while the halo accretion material makes up around 60 per cent. The halo accretion contribution remains at this relative contribution and even drops slightly in the outskirts, but the DM material deposited via the smooth accretion sub-channel becomes increasingly important at larger cluster radii, reaching 50 per cent of the mass budget at $R_{178}$. This ordering of the radii at which the material in each assembly (sub-)channel becomes significant to the radial profile (early assembly, halo accretion then smooth accretion) reflects the ordering of the $\langle \varepsilon \rangle$ and $\langle j \rangle$ seen in Section \ref{sec:results_assembly_channels}.

\subsubsection{Morphological shapes}\label{sec:results_morphological_shapes}

The morphological shape profiles of the BCG+ICL and main halo material, divided into assembly (sub-)channels, are measured. The projected mass distribution is taken along three orthogonal axes for each cluster, resulting in 33 projections in total, which are treated as independent. The shape is characterised by the axis ratio $p=b/a$, where $a$ and $b$ are the semi-major and semi-minor axes respectively of the equivalent ellipse describing the mass distribution in an annulus. This scheme is the same as in \cite{Fernandez_2026}, measuring the shape in 10 logarithmically-spaced radial bins from $a=0.01R_{178}$ to $a=R_{178}$ using the mass-weighted moment-of-inertia tensor (also known as the shape tensor) of the projected mass distribution, with the centre fixed as the cluster halo centre. For each $p$ measurement of assembly (sub-)channel material, we impose that a minimum of 10 per cent of the BCG+ICL (or main halo) material must originate from that (sub-)channel.

The top two panels of Fig.~\ref{fig:shapes} display the morphological shape profiles of the BCG+ICL (left) and main halo (middle). The crosses and shaded region represent the sample mean and bootstrapped error on the mean respectively. If less than half of the sample has a $p$ measurement for a given point, that point is not displayed. This is the case for the early assembly profile in the outer regions and the smooth accretion profile in the inner regions.

The total BCG+ICL shape profile becomes more elliptical at larger $a$, decreasing from $p\sim0.7$ to $p\sim0.5$, which is consistent with observations that the BCG is rounder than the ICL \citep{Kolcu_2026}. The main halo profile is flatter, remaining at $p\sim0.65$ and becoming rounder in the cluster outskirts. The early assembly profile is substantially rounder than the late assembly profile for both the BCG+ICL and main halo. However, differences between the mean profiles do not necessarily reflect systematic differences within individual clusters, given we find a substantial dispersion in shape profiles across the sample. To investigate whether this difference persists on a cluster-by-cluster basis, the bottom panels show $\Delta p$, the difference in $p$ between two profiles within a given sample. $\Delta p$ between the early and late assembly is shown in grey -- we indeed see that typically $\Delta p\sim0.15$ for both the BCG+ICL and main halo.

The halo accretion and smooth accretion shape profiles for the main halo are also shown in the top middle panel. The smooth accretion profile is very round ($p\sim0.75$), whilst the halo accretion profile becomes progressively rounder towards larger $a$. The difference between these two profiles is shown in purple on the bottom middle panel, with $\Delta p$ decreasing steadily as $a$ increases.

Finally, we show $\Delta p$ for each (sub-)channel between the main halo and BCG+ICL profiles on the bottom right panel. Except for the innermost $a$, the main halo profile is always rounder than the BCG+ICL profile, both in total and for the early and late assembly profiles separately. $\Delta p$ grows larger as $a$ increases, driven by the main halo profile becoming more circular while the BCG+ICL profile remains flat. Since the DM has the additional smooth accretion assembly sub-channel, we also plot $\Delta p$ between the halo accretion profile for the main halo and the late assembly profile for the BCG+ICL (which is the same as the halo accretion profile). Although $\Delta p$ is smaller compared to $\Delta p$ for the late assembly channel, the shape profile offset between the BCG+ICL and main halo is not reconciled.

The connection between the morphological shape and $\langle \varepsilon \rangle$ or $\langle j \rangle$ of the material associated with each assembly channel is not straightforward. For the main halo, the smooth accretion and early assembly material are the highest and lowest $\langle \varepsilon \rangle$ components respectively, and yet their shape profiles are both substantially rounder than the halo accretion material. Other factors, such as accretion geometry \citep{Martin_2026b} and the process of phase-mixing, may also be influencing the morphological shape profiles. We develop these ideas further in Section \ref{sec:discussion_morphological_shapes}.

\section{Discussion}\label{sec:discussion}

A strong candidate for the mechanism to explain the offset between the orbital energy/angular momentum of the (BCG+)ICL stars and DM is the differential stripping of stars and DM from haloes. The loosely-bound DM is liberated rapidly upon cluster infall and retains its initial orbital properties, whereas the tightly-bound stars remain in haloes until stripped on lower energy and angular momentum orbits, whilst these haloes undergo orbital decay through processes such as dynamical friction. We previously demonstrated in \citetalias{Martin_2026a} that this mechanism leads to an offset between the orbital energies of the stars and DM in idealised merger simulations. When convolved with a realistic satellite infall population this reproduces an offset similar to that found by \citetalias{Butler_2025} and other cosmological simulations. Here we show that, in a full cosmological context, this mechanism is responsible for much of the offset. Fig.~\ref{fig:se_per_merger_bcg_icl} shows that major mergers, for which infalling haloes experience the strongest dynamical decay, lead to a substantial deficit in $\langle \varepsilon \rangle$ for the stars compared to the DM.

To summarise our results: we find that the assembly time of material is important for the orbital and spatial distributions at $z\approx0$, with early assembly material being on lower energy/angular momentum orbits, more centrally-concentrated, and morphologically rounder than the late assembly material. Phase space offsets between the (BCG+)ICL stars and DM are observed for the early and late assembly material separately. Merger time and merger mass ratio impact the degree of differential stripping of stars and DM from satellites, being stronger for earlier and closer mass ratio mergers. We also show that mergers alone cannot fully account for the phase space offsets between the (BCG+)ICL stars and DM -- the presence of DM material accreted through the smooth accretion sub-channel, which preferentially occupies high energy/angular momentum orbits compared to the halo accretion sub-channel material, contributes significantly as well.

\subsection{Assembly time \& cluster evolution}\label{sec:discussion_assembly_time}

Cluster assembly in cosmological simulations proceeds by the continuous accretion of haloes and surrounding material in a self-consistent, hierarchical fashion. In \citetalias{Martin_2026a}, this process was approximated by performing controlled merger simulations across a range of merger mass ratios, then convolving the results with realistic satellite infall populations, such that each merger effectively occurred at $z=0.6$. For each merger, the orbital properties of the stripped stars and DM were taken after 6\,Gyr to estimate the $\langle \varepsilon_\textrm{ratio} \rangle$. The justification for this was that in an adiabatically-evolving system, the $\langle \varepsilon_\textrm{ratio} \rangle$ of the material liberated from individual mergers should remain approximately constant, since the orbital action is conserved \citep{Binney_Tremaine_2008}.

However, the \textit{absolute} $\langle \varepsilon \rangle$ of the stars and DM liberated from a given merger is dependent on the merger time, and the progression of the tidal stripping of material (particularly for stars) is similarly dependent on how long the satellite has spent in the cluster. The range of merger times provided by the full cosmological context of the simulation in this work is therefore necessary to produce a more realistic $\langle \varepsilon \rangle$, since the halo accretion sub-channel material is a mix of material from these mergers.

If non-adiabatic processes are significant during cluster assembly, then material accreted at different times could be redistributed to reflect the state of the cluster at $z\approx0$, gradually erasing any memory of when it joined the cluster. Instead, the early and late assembly material still occupy distinct regions of phase space at $z\approx0$ (Figs.~\ref{fig:se_am_assembly_channel} and \ref{fig:radial_contributions}), suggesting that cluster evolution since $z\approx2$ has been close enough to adiabatic for this memory to survive. The assumption of adiabatic cluster evolution underpinning both \citetalias{Martin_2026a} and the analysis in this work therefore holds. It also shows that the orbital offset between the stars and DM is largely set at the time each merger’s material is stripped and then preserved, so the $z\approx0$ value of $\langle \varepsilon_\textrm{ratio} \rangle$ reflects the cluster’s assembly history: which mergers occurred, when, and how far their stripping had progressed by $z\approx0$.

We also acknowledge the effect of pseudo-evolution on cluster growth. This refers to the apparent evolution of halo mass due to the redshift evolution of the reference density used in the halo definition, which in the \textsc{AdaptaHOP} structure finder used in this work is the total matter density of the entire simulation box. This effect is most prominent for lower mass haloes, but can still account for a significant fraction of cluster mass evolution \citep{Diemer_2013}. This apparent growth is difficult to separate from genuine cluster growth through mergers and especially smooth accretion.

To estimate the contribution of pseudo-evolution to the late assembly mass budget, we assess the DM material in a shell around the cluster centre at $z\approx2$. The inner radius of this shell is the cluster $R_{178}$ at $z\approx2$, and the outer radius is the co-moving cluster $R_{178}$ at $z\approx0$. The volume within this shell therefore is equal to the co-moving volume the cluster occupies at $z\approx0$, minus the volume (in physical units) of the cluster $R_{178}$ at $z\approx2$. We find that $<3$ per cent of the late assembly DM material is located within this shell at $z\approx2$, therefore pseudo-evolution does not make a substantial contribution to the cluster growth after $z\approx2$.

\subsection{Smooth accretion}\label{sec:discussion_smooth_accretion}

The presence of the smooth accretion sub-channel is predicted in a $\Lambda$CDM Universe -- using the Millennium \citep{Springel_2005} and Millennium II \citep{Boyland-Kolchin_2009} cosmological N-body simulations, \cite{Genel_2010} find that the DM mass fraction contribution of all mergers to a $z=0$ host halo saturates at $\approx60$ per cent when increasingly-unequal mass ratio mergers are resolved, such that $\approx40$ per cent of the DM mass is accreted in a genuinely smooth fashion, never belonging to a halo before that point. With careful construction of merger trees and fitting of the merger rate, \cite{Genel_2010} still find this approximate result when the mass contribution from mergers is instead obtained by integrating over the merger rate.

Other fitting formulae for the merger rate, such as the extended Press-Schechter (EPS) model, by construction predict no smooth accretion. This "extra mass" required is instead accounted for by higher merger rates, which can be within both the closer mass ratio merger regime and the extremely-unequal mass ratio merger regime. The latter often originates from unrealistic extrapolation of merger rates to unresolved merger ratios, rather than the mass contribution plateauing. This plateau also means the mass contribution from extremely low mass haloes should be vanishingly small before reaching a minimum halo mass set by the free-streaming mass of the DM model. The existence of a smooth accretion channel in our work is therefore not simply a product of unresolved haloes due to the limited resolution of the simulation -- see \cite{Genel_2010} for further discussion.

Under the assumption that the average ratio between the DM mass in halo accretion and smooth accretion sub-channels in the late assembly channel holds for the early assembly, our smooth accretion channel accounts for $\approx30$ per cent of the total DM in the cluster. Although the implementation of various aspects of galaxy physics (e.g. star formation, mechanical feedback) may differ between cosmological simulations, they all model DM as a collisionless component evolving under gravity alone. Since halo assembly is driven predominantly by gravity, we therefore expect this smooth accretion channel to be a persistent feature across models. However, significant differences may arise in the identification of this material, due to the wide range of structure-finder algorithms used to identify haloes, and any further methodological choices to delineate between material assigned to the halo accretion and smooth accretion sub-channels. Ultimately, this reflects the nature of DM haloes, being embedded in larger cosmological structures and often still accreting material themselves as they merge with the cluster, meaning there is no clear boundary between halo and smooth accretion.

In \citetalias{Martin_2026a}, we did not model a smooth accretion assembly channel; instead, all the accreted DM material enters the cluster in resolved haloes. Nevertheless, in \citetalias{Martin_2026a} we obtained an envelope of predicted $\langle \varepsilon_\textrm{ratio} \rangle$ values which overlap with the measured $\langle \varepsilon_\textrm{ratio} \rangle$ values in \textsc{Horizon-AGN}. We suggest this is possible because dynamically, the DM we attribute to the smooth accretion sub-channel acts very similarly to any DM stripped from haloes immediately or shortly after infall, which occurs for the bulk of the infalling halo's DM material following the first pericentric passage \citep[][\citetalias{Martin_2026a}]{Smith_2016b, Martin_2024}. Throughout the time the infallen halo undergoes dynamical processes that change its orbit, the smooth accretion and quickly-stripped DM material is not subject to these processes, and thus their orbits should not change substantially from infall. The smooth accretion material occupies the highest $\langle \varepsilon \rangle$ and $\langle j \rangle$ of any (sub-)channel, implying the particles have lost little or no energy and angular momentum. Future studies into the BCG+ICL and DM halo connection using controlled simulations or semi-analytical models should explicitly include this smooth accretion channel. Since this material is unaffected by dynamical friction or stripping this would be straightforward to implement, and could be incorporated by placing a fraction of the accreted DM directly on orbits drawn from haloes measured in cosmological simulations (e.g. \citealt{Wetzel_2011, Jiang_2015}).

\subsection{Morphological shapes}\label{sec:discussion_morphological_shapes}

The morphological shape of the BCG+ICL and main halo material at a given radius are set by shape of the dominant assembly channel for material at that radius. The overall shape profiles are therefore connected to the orbital properties of material from different assembly channels; we show that the higher energy/angular momentum late assembly material dominates the outer cluster regions of both the BCG+ICL and main halo, and conversely the lower energy/angular momentum early assembly material greatly contributes to the inner cluster regions. For the DM there is the additional smooth accretion sub-channel, with this material becoming increasingly important at large radii and influencing the overall main halo profile to a rounder shape (although the halo accretion profile also becomes more circular in the cluster outskirts).

The underlying mechanisms that determine the shape profiles of the material from each assembly channel are less clearly related to orbital energies and angular momenta. To first order, the shape of the liberated material is set by the accretion geometry and time, which broadly leads to the correspondence between the BCG+ICL and main halo shape profiles \citep{Martin_2026b}. The details of this are more complicated however -- material is continuously stripped from satellites, and thus deposited on orbits with a range of spatial and temporal offsets. The material stripped from the same satellite is therefore related, but the material in a given annulus is a overlapping mix of material from many satellites, and material from the smooth accretion sub-channel, all of which have evolving orbits. Suffice to say, the orbital properties of the infalling populations do not have a direct mapping to the resulting shape profiles.

A possible explanation for the similarity in the shapes of the early assembly and smooth accretion material is the degree of phase-mixing. This is the process by which distinct remnants of tidal stripping and liberation of material from an infalling object, such as tidal streams and shells, are dissipated by spreading out in phase space.\footnote{The process of phase-mixing preserves the phase-space density of the stripped material, but due to small differences in initial orbits of individual particles, distinct objects become increasingly less obvious in configuration space over time.} Since the process of phase-mixing begins once material has been liberated, we suggest that the material that has spent longer as part of the cluster halo has longer to phase-mix and result in a rounder component on average. For the early assembly material, this extended phase-mixing time is due to earlier accretion onto the cluster, and for the smooth accretion is due to being unbound to any (resolvable) halo upon infall.

\section{Conclusions}\label{sec:conclusions}

In this study, we have investigated the orbital properties and spatial distributions of the BCG+ICL stars and the main halo DM at $z\approx0$, in 11 clusters from the \textsc{Horizon-AGN} cosmological hydrodynamic simulation. Our results are summarised below:

\begin{enumerate}
    \item We categorise the $z\approx0$ cluster stellar and DM material into assembly channels. Compared to the late assembly material ($z\lesssim2$), the early assembly stars and DM ($z\gtrsim2$) preferentially occupy lower energy and angular momentum orbits, are more centrally-concentrated, and are morphologically rounder.
    \item For the early and late assembly channels separately, the associated stellar material preferentially occupies lower energy and angular momentum orbits than the DM from the same assembly channel. The phase-space offset between the (BCG+)ICL and the DM is therefore not only a consequence of the two components being assembled through different channels. It arises within each channel, and is preserved as the cluster grows.
    \item Merger time and merger mass ratio impact the degree of differential stripping of stars and DM from satellites by $z\approx0$, being stronger for earlier and closer mass ratio mergers. The orbits of more equal-mass satellites decay through dynamical friction, and the satellite material is stripped in an outside-in fashion. The halo outskirts are dominated by DM and thus get deposited at larger radii/higher orbital energy and angular momentum orbits, whilst stars are stripped later and preferentially end up at smaller radii/lower energy and angular momentum orbits. Additionally, earlier mergers have had longer to complete this process, so a larger share of their stars has been stripped deep in the potential.
    \item The DM material has a significant smooth accretion sub-channel, representing the material infallen as part of larger cosmic structures and the outskirts of merging haloes, for which there is no stellar counterpart. This material preferentially occupies very high energy and angular momentum orbits, thus has not undergone substantial orbital decay after infall. The material in this channel contributes substantially to the (BCG+)ICL and DM phase space offsets.
\end{enumerate}

Despite all these complexities, the orbital energy ratio between the BCG+ICL stars and main halo DM has a relatively small dispersion between individual galaxy clusters. This is promising for using the (BCG+)ICL as a DM tracer, but our sample is too small to properly assess the impact of dynamical state, assembly history and cluster parameters on these ratios. This will be explored in an upcoming study using a much larger sample of clusters.

\section*{Acknowledgements}

M.~B, A.~F and N.~A.~H gratefully acknowledge support from the Leverhulme Trust through a Research Leadership Award. N.~A.~H, G.~M, and F.~R.~P acknowledge support from the UK STFC under grant ST/X000982/1. H.~J.~B acknowledges support from the UK Science and Technology Facilities Council (STFC) under grant ST/Y509437/1. This research was supported by the International Space Science Institute (ISSI) in Bern, through ISSI International Team project \#23-577. We also thank members of the NottICL group for helpful discussions and comments.

The \textsc{Horizon-AGN} simulation was granted access to the HPC resources of CINES under allocations 2013047012, 2014047012 and 2015047012 made by GENCI, and made use of the Infinity cluster, hosted by the Institut d’Astrophysique de Paris. We warmly thank S. Rouberol for running it smoothly.

\section*{Data Availability}

The raw data products of the {\sc Horizon-AGN} simulation are available upon reasonable request through the collaboration’s website: \url{https:// www.horizon-simulation.org/}. The data products generated from this work are available upon reasonable request from the corresponding author.


\bibliographystyle{mnras}
\bibliography{paper} 

\begin{appendix}

\section{Additional Results}\label{sec:appendix_additional_results}

\begin{figure*}
    \centering
    \begin{minipage}{\columnwidth}
        \includegraphics[width=\columnwidth]{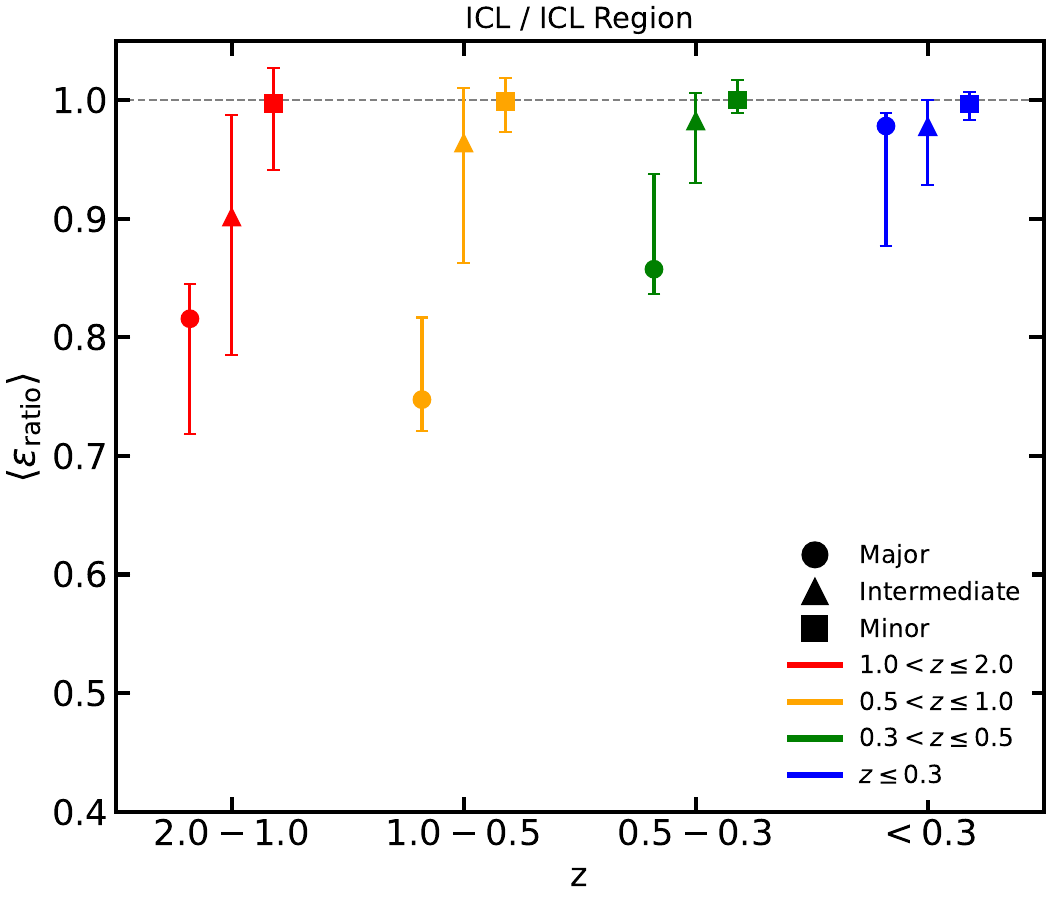}
        \caption{The same as Fig.~\ref{fig:se_per_merger_bcg_icl} but for exclusively the ICL rather than the combined BCG+ICL.}
        \label{fig:se_per_merger_icl}
    \end{minipage}

    \vspace{0.5cm}

    \includegraphics[width=\textwidth]{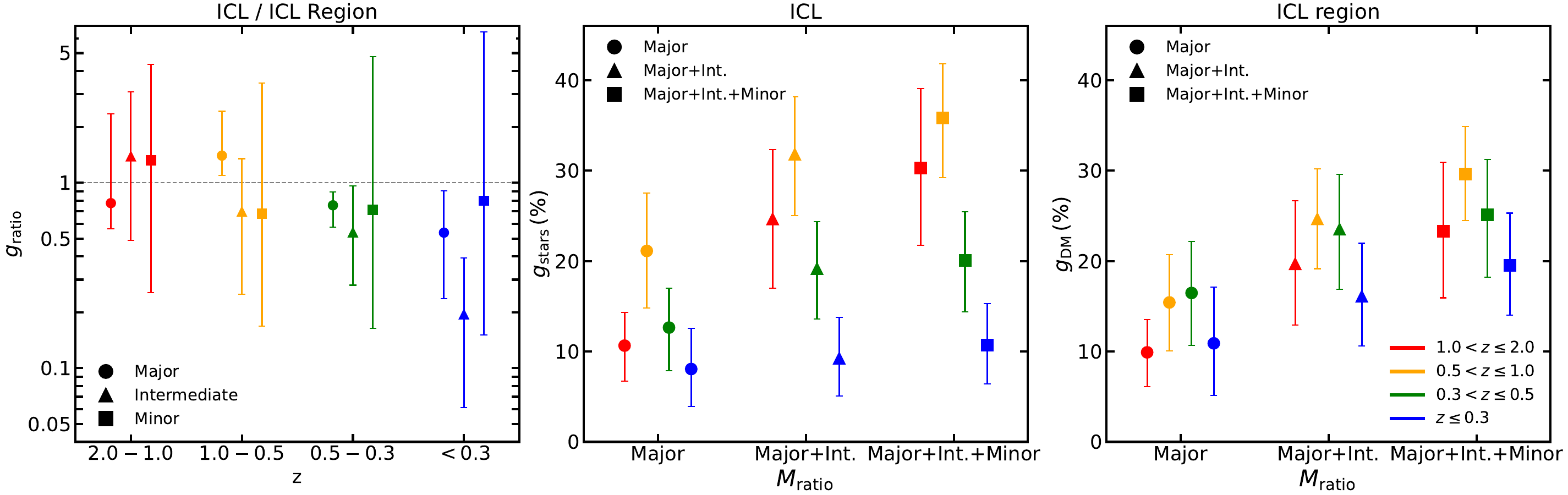}
    \caption{The same as Fig.~\ref{fig:merger_contributions_bcg_icl} but for exclusively the ICL rather than the combined BCG+ICL.}
    \label{fig:merger_contributions_icl}
\end{figure*}

We present the equivalent versions of Figures \ref{fig:se_per_merger_bcg_icl} \& \ref{fig:merger_contributions_bcg_icl} for the ICL in Figures \ref{fig:se_per_merger_icl} \& \ref{fig:merger_contributions_icl} respectively.

\end{appendix}

\label{lastpage}
\end{document}